\documentclass[11pt,a4paper]{article} \usepackage{jheppub}

\newcommand{\as}{\alpha_{\mathrm{s}}}

\newcommand{\misq}{m_i^2}
\newcommand{\PijQ}{P_{ij} \cdot Q}
\newcommand{\Pijsq}{P_{ij}^2}
\newcommand{\Qsq}{Q^2}
\newcommand{\beq}{\begin{equation}}
\newcommand{\eeq}{\end{equation}}
\newcommand{\bea}{\begin{eqnarray}}
\newcommand{\eea}{\end{eqnarray}}
\newbox\charbox
\newbox\slabox
\def\s#1{{      % Feynman slash
        \setbox\charbox=\hbox{$#1$}
        \setbox\slabox=\hbox{$/$}
        \dimen\charbox=\ht\slabox
        \advance\dimen\charbox by -\dp\slabox
        \advance\dimen\charbox by -\ht\charbox
        \advance\dimen\charbox by \dp\charbox
        \divide\dimen\charbox by 2
        \raise-\dimen\charbox\hbox to \wd\charbox{\hss/\hss}
        \llap{$#1$}
}}

\title{Complete Nagy-Soper subtraction for next-to-leading order
  calculations in QCD}

\author[]{G. Bevilacqua, M. Czakon, M. Kubocz and M. Worek} 
\affiliation[]{Institut f\"ur Theoretische Teilchenphysik und Kosmologie,
RWTH Aachen University, D-52056 Aachen, Germany} 
\emailAdd{bevilacqua@physik.rwth-aachen.de}
\emailAdd{mczakon@physik.rwth-aachen.de}
\emailAdd{kubocz@physik.rwth-aachen.de} 
\emailAdd{worek@physik.rwth-aachen.de} 

\abstract{We extend the \textsc{Helac-Dipoles} package with the
  implementation of a new subtraction formalism, first introduced by
  Nagy and Soper in the formulation of an improved parton shower.  We
  discuss a systematic, semi-numerical  approach for the evaluation of
  the integrated subtraction  terms for both massless and massive
  partons,  which provides the missing ingredient for a complete
  implementation. In consequence, the new scheme can now be used as
  part of a complete NLO QCD calculation for processes with  arbitrary
  parton masses and multiplicities. We assess its overall performance
  through a detailed comparison with results based on Catani-Seymour
  subtraction. The importance of random polarization and color sampling 
  of the external partons is also examined.}

\dedicated{\rm TTK-13-20}

\begin{document} 
\maketitle
\flushbottom

%
%-------------%-------------%-------------%-------------

\section{Introduction}

Recent years have seen a tremendous progress in next-to-leading order (NLO)
calculations \cite{Bredenstein:2009aj,Bevilacqua:2009zn,
Bredenstein:2010rs,Bevilacqua:2010ve,Bevilacqua:2011aa,Worek:2011rd,
Berger:2009zg,KeithEllis:2009bu,Berger:2010zx,Bern:2013gka,Denner:2010jp,
Bevilacqua:2010qb,Melia:2011dw,Greiner:2012im,Denner:2012yc,
Bevilacqua:2013taa,Becker:2011vg,Bern:2011ep,Badger:2012pf,
Greiner:2011mp,Bevilacqua:2012em,Campanario:2011ud,Campanario:2013mga,
vanDeurzen:2013xla,Gehrmann:2013bga,Denner:2012dz,Cullen:2013saa}. 
Most it was due to the development of new
methods \cite{Bern:1994zx,Britto:2004nc,Ossola:2006us,Giele:2008ve,
Cullen:2011ac,Cascioli:2011va,Actis:2012qn,Bevilacqua:2011xh,vanHameren:2009dr} 
and the perfectioning of traditional approaches 
\cite{Binoth:2005ff,Denner:2005nn,Arnold:2008rz,Cullen:2011kv} to the
calculation of virtual corrections. In practical applications,
however, most computational time is spent on the evaluation of real
radiation corrections. The latter are not considered a technical
obstacle, because of subtraction schemes, which allow to reduce
the problem to the application of ordinary Monte Carlo methods and
were already introduced in the 90's \cite{Frixione:1995ms,Catani:1996vz}. 
This paper is concerned with another
subtraction scheme, named after Z. Nagy and D. Soper, and based on a
new concept of a parton shower presented in 
\cite{Nagy:2007ty,Nagy:2008ns,Nagy:2008eq}. The ideas of the
latter publication have already been partly exploited in a series of
papers \cite{Chung:2010fx,Chung:2012rq,Robens:2013wga}. The present 
publication is meant to complete the
construction for all cases of practical interest.

Since the Nagy-Soper subtraction scheme is based on a parton shower,
it should be obvious that our main motivation is to provide a
framework for simple matching between a fixed order calculation and
the new parton shower. This topic will be tackled in future
publications. Here, we wish to solve the problem of the integration of
subtraction terms over the unresolved phase space. This is the
non-trivial part of any subtraction scheme. Contrary to the usual
practice at NLO, we will not perform involved analytic
integrations. After a suitable parameterization, we will perform the
integrations numerically much in the spirit of most recent NNLO
methods \cite{Czakon:2010td}. This will allow us to cover both massless
 and massive
cases with similar effort. At the same time, we will argue that the
cost of the numerical integrations in applications may be
neglected. This semi-numerical approach is what sets us apart from the
publications \cite{Chung:2010fx,Chung:2012rq,Robens:2013wga}.

In order to exploit the available tools as much as possible, we will
integrate the new subtraction scheme within the \textsc{Helac-Dipoles}
framework \cite{Czakon:2009ss}, 
which already provides most of the ingredients, {\it i.e.} the
color and spin correlated tree-level matrix elements and the Monte
Carlo integration system. We will also demonstrate the capabilities of
our system on a few challenging examples.

This paper is organized as follows: in the next section, section 2, we will
review the concept of subtraction schemes and point out the
differences between the most used Catani-Seymour subtraction 
\cite{Catani:1996vz,Catani:2002hc} and
the Nagy-Soper subtraction; subsequently, we will present the details
of the Nagy-Soper approach, {\it i.e.} momentum mappings, splitting and soft
functions and phase space factorizations; in section 4, we will
explain our approach to the integration of subtraction terms, while in
section 5 we will discuss the implementation inside 
\textsc{Helac-Dipoles}; section 6 will contain numerical comparisons between
our implementation of the Nagy-Soper and Catani-Seymour subtractions, 
together with an assessment of overall performance of  
random polarization and color sampling. 

%-------------%-------------%-------------%-------------
%

%
%-------------%-------------%-------------%-------------

\section{General framework of subtraction at NLO QCD}
\label{sec:general_framework}
Let us consider a generic process involving $m+1$ external QCD partons with 
momenta $p_a+p_b \to p_1+\cdots + p_{m-1}$: we outline the method for the most 
general case where an unresolved parton can be radiated off either the initial 
or the final state, typical of a hadron collider. The same approach applies to 
other cases as well, such as $e^+e^-$ colliders, with some  conceptual 
simplifications that will be clear in the following. 
The inclusive cross section, at NLO QCD accuracy, reads
\bea
\sigma_{\rm NLO} & = & \int_{m} d\Phi_{m} \,\, \mathcal{A}^{B}(\{p\}_{m}) 
\,\, F_{m} \\ 
& + & \int_{m+1} d\Phi_{m+1} \,\, \mathcal{A}^{R}(\{p\}_{m+1}) \,\, F_{m+1} \\
& + &  \int_{m} d\Phi_{m} \,\, \mathcal{A}^{V}(\{p\}_{m}) \,\, F_{m}  \\
& + &  \int_0^1 dx \int_{m} d\Phi_{m}(x) \,\, \mathcal{A}^{C}(x,\{p\}_m) 
\,\, F_m
\eea
where we have neglected (in the remainder of this section as well) the global
flux, statistical, spin and color averaging factors, moreover
\beq
\mathcal{A}^{B} \equiv \vert \mathcal{M}^{\mbox{\scriptsize{Born}}} 
\vert^2 \,,\;\;\;\;\:  \mathcal{A}^{R} \equiv \vert 
\mathcal{M}^{\mbox{\scriptsize{Real}}} \vert^2  \,,\;\;\; \:\: 
\mathcal{A}^{V} \equiv 2\,\Re \left[ \mathcal{M}^{\mbox{\scriptsize{Born}}} 
\, ( \mathcal{M}^{\mbox{\scriptsize{1-Loop}}} )^* \right]\,,
\eeq
where $\mathcal{M}^{\mbox{\scriptsize{Born}}}$, 
$\mathcal{M}^{\mbox{\scriptsize{1-Loop}}}$, 
$\mathcal{M}^{\mbox{\scriptsize{Real}}}$ represent the Born, one-loop and 
real-emission matrix elements respectively. We denote with $d\Phi_{m}$ the 
integration measure for a $m$-parton phase space and with $F_{m}$ the jet 
function which shapes the kinematically accessible regions of it.

When the kinematics of the final state is characterized by $m$ well separated 
hard jets, the Born contribution is finite, whereas the virtual and the 
real-emission terms are individually divergent due to the presence of 
singularities of various nature: ultraviolet for $\mathcal{A}^V$, and 
soft/collinear for both $\mathcal{A}^V$ and $\mathcal{A}^R$. As it is well
 known, when using an infrared-safe definition of partonic jets, all soft and 
collinear divergencies that affect the virtual and real corrections cancel 
in the sum, with one notable exception: the singularities arising from the 
emission of nearly-collinear partons off the initial state. In fact, the 
latter are reabsorbed into a re-definition of the parton distribution 
functions (PDFs) that is achieved by introducing suitable 
\textit{collinear counterterms}, $\mathcal{A}^C$. In the 
$\overline{\mbox{MS}}$ scheme with $d=4-2\epsilon$ dimensions, 
these counterterms read as follows,
\bea
& & \int_0^1 dx \int_{m} d\Phi_{m}(x) \,\, \mathcal{A}^{C}(x,\{p\}_m) 
\,\,F_{m} = \\
& = & \frac{\as}{2\pi}\,\frac{1}{\Gamma(1-\epsilon)}
\,\sum_{k}\int_0^1 dx \int_m d\Phi_{m}(x) \,\, 
\mathcal{A}^B_{ak}(xp_a,p_b)\,
\frac{1}{\epsilon}\left(\frac{4\pi\mu_R^2}{\mu_F^2}\right)^\epsilon P_{ak}(x) 
\,\,F_{m}   \label{eq:AltarelliParisi1} \\
& + & \frac{\as}{2\pi}\,\frac{1}{\Gamma(1-\epsilon)}\,\sum_{k}\int_0^1 dx 
\int_m d\Phi_{m}(x) \,\, \mathcal{A}^B_{bk}(p_a,xp_b)\,
\frac{1}{\epsilon}\left(\frac{4\pi\mu_R^2}{\mu_F^2}\right)^\epsilon 
P_{bk}(x) \,\,F_{m} \;. \label{eq:AltarelliParisi2}
\eea
Here $P_{ab}(x)$ are the Altarelli-Parisi kernels in four dimensions 
\cite{Altarelli:1977zs}, while $\mu_R$, $\mu_F$ denote the renormalization 
and factorization scale respectively. We use the symbolic notation 
"$d\Phi_m(x)$" to underline that the kinematically available phase space  
depends upon the value of the integration variable $x$, since the
square of the
center-of-mass energy of the partonic system (neglecting parton masses) 
is $\hat{s}(x)=2 x p_a \cdot p_b$. Also, it should be clear that in all 
cases where no initial-state QCD radiation is possible, as for example 
in the case of $e^+e^-$ collisions, no contribution from the collinear 
counterterms is necessary.

As a consequence of the cancellations, and after ultraviolet renormalization 
is performed, the inclusive cross section $\sigma_{\rm NLO}$ 
is a finite quantity. 
However, the individual pieces still suffer from soft and collinear 
divergencies and cannot be integrated numerically in four dimensions. 
A possible way to overcome this issue is to regularize the integrands 
by adding and subtracting local counterterms, $\mathcal{A}^D$, designed 
to match the singular structure in the soft and collinear limits:
\bea
 \sigma_{\rm NLO} & = & \int_{m} d\Phi_{m} \,\, \mathcal{A}^{B}(\{p\}_{m}) 
\,\, F_{m} \\
& + & \int_{m+1} d\Phi_{m+1} \, \left[ \mathcal{A}^{R}(\{p\}_{m+1}) 
\,\, F_{m+1} - \mathcal{A}^{D}(\{p\}_{m+1}) \,\,F_{m}  \right] 
\label{eq:real} \\
& + & \int_0^1 dx \int_m d\Phi_{m}(x) \left[ \, \delta(1-x) 
\left( \mathcal{A}^V(\{p\}_m) + \int_1 \mathcal{A}^D(\{p\}_{m+1}) \right) 
\right. \nonumber \\ 
& & \left. + \,\, \mathcal{A}^C(x,\{p\}_m) \, \right] \, F_{m} .  
\label{eq:virtual} 
\eea
As in the case of the real-emission terms, the local counterterms 
$\mathcal{A}^{D}$ are defined on the $(m+1)$-parton phase space, 
denoted $\{p\}_{m+1}$. They are subtracted from $\mathcal{A}^R$ and 
added back to $\mathcal{A}^V$ after integration over the phase space 
of the unresolved parton. The outlined \textit{subtraction method} 
makes the integrals (\ref{eq:real}) and (\ref{eq:virtual}) individually  
convergent and thus well suited for a Monte Carlo integration.

The construction of the local counterterms is inspired by the well known 
property of the universal factorization of QCD amplitudes in the soft and 
collinear limits. Schematically, the singular structure 
of a $(m+1)$-parton squared amplitude for two partons $p_i$ and $p_j$ 
becoming collinear can be approximated as follows:
\beq
\langle \mathcal{M}(\{p\}_{m+1}) \vert \mathcal{M}(\{p\}_{m+1}) 
\rangle _{sing}
  \,  \approx \, \langle \mathcal{M}(\{\overline{p}\}_{m}^{(ij)}) 
\vert \left( \mathbf{V}_{ij}^\dagger \cdot \mathbf{V}_{ij} \right) 
\vert  \mathcal{M}(\{\overline{p}\}_{m}^{(ij)}) \rangle \nonumber \;.
\eeq
Here $\vert\mathcal{M}(\{\overline{p}\}_{m})\rangle$ is an amplitude 
describing $m$ on-shell external partons and $\mathbf{V}$ is an operator 
acting on the spin part of the amplitude, while $\{\overline{p}\}_{m}^{(ij)}$ 
represents the $m$-parton kinematics to which $\{p\}_{m+1}$ reduces in the 
limit where partons $p_i$ and $p_j$ are strictly collinear: the two splitting 
partons merge to form a new on-shell parton, $\overline{p}_i \equiv
p_i \pm p_j$ 
(the sign depends whether the splitting occurs in the final or in the initial 
state), while the remaining momenta are left unchanged. In other words, using 
our standard notation, in the regions of the phase space where the pair 
$(p_i,p_j)$ is nearly collinear, the structure of the real-emission 
contribution simplifies to 
\beq
\mathcal{A}^{R}(\{p\}_{m+1}) \, \approx \,  
\mathcal{A}^{B}(\{\overline{p}\}_{m}^{(ij)}) \otimes 
\mathcal{C}^{(ij)}(\overline{p}_i;p_i,p_j) \;.
\eeq
Namely, it reduces to the product of a finite Born squared amplitude 
times a divergent, collinear splitting kernel $\mathcal{C}^{ij}$ associated 
with the splitting $\overline{p}_i \to p_i+p_j$. The symbol $\otimes$ denotes 
here spin correlations. On the other hand, when a parton $p_j$ becomes soft,  
factorization takes the form
\beq
\langle \mathcal{M}(\{p\}_{m+1}) \vert \mathcal{M}(\{p\}_{m+1}) 
\rangle _{sing}
  \,  \approx \, \sum_{k \ne j} \langle \mathcal{M}
(\{\overline{p}\}_{m}^{(j)}) \vert \left( \mathbf{T}_{i} 
\cdot \mathbf{T}_{k} \right) \vert  
\mathcal{M}(\{\overline{p}\}_{m}^{(j)}) \rangle \nonumber \;,
\eeq
where $\mathbf{T}$ is an operator acting on the color part of the amplitude 
and $\{\overline{p}\}_{m}^{(j)}$ is the soft limit of the kinematical 
configuration $\{p\}_{m+1}$, where $\overline{p}_i \equiv p_i$ and the 
remaining momenta are left unchanged. Using our notation, in the nearly-soft 
limit where $p_j \to 0$, one can write
\beq
\mathcal{A}^{R}(\{p\}_{m+1}) \, \approx \, \sum_{k \ne j} 
\mathcal{A}^{B}(\{\overline{p}\}_{m}^{(j)}) \otimes \mathcal{S}^{(kj)}
(\overline{p}_i,\overline{p}_k;p_i,p_k,p_j) \;,
\eeq
and the factorization is expressed in terms of $m$ soft splitting kernels 
$\mathcal{S}^{(kj)}$, one for each external parton. The symbol $\otimes$ 
denotes here color correlations.

The outlined factorization properties suggest a general rule for constructing 
the local counterterms  within the subtraction method. First, one needs to 
define a complete set of transformations which map the original phase 
space $\{p\}_{m+1}$ into a new space of $m$ on-shell partons, 
called $\{\tilde{p}\}_{m}$:
\beq
\{p\}_{m+1} \to \{\tilde{p}\}_{m}^{(\ell)}  \;\;\;,\;\;\; \ell=\{1,...,N\} \;.
\eeq
By complete, we mean that for each pair $(p_i,p_j)$ becoming collinear, 
there is at least one mapping $\{\tilde{p}\}_{m}^{(\ell)}$ that smoothly 
approaches the singular kinematical configuration $\{\overline{p}\}_{m}^{(ij)}$. 
Similarly, for each parton $p_j$ becoming soft, there is at least one $\ell$ 
such that $\{\tilde{p}\}_{m}^{(\ell)} \to \{\overline{p}\}_{m}^{(j)}$. 

Second, one needs to define a set of splitting functions 
$\mathcal{D}^{(\ell)}(\{\tilde{p}\}_{m},\{p\}_{m+1})$, that match the 
behavior of the soft and collinear kernels in the singular limits. 
Then, the local counterterms take the general form:
\bea
\mathcal{A}^{D}(\{p\}_{m+1}) = \sum_{\ell=1}^{N} 
\mathcal{A}^B(\{\tilde{p}\}_{m}^{(\ell)}) \otimes 
\mathcal{D}^{(\ell)}(\{\tilde{p}\}_{m}^{(\ell)},\{p\}_{m+1}) \;.
\eea
There is of course a certain freedom in defining both mappings and 
splitting functions away from the singular limits, each choice leading 
to a different subtraction scheme. Perhaps the most widespread version 
is the Catani-Seymour (CS) scheme \cite{Catani:1996vz,Catani:2002hc}, 
where one has 
\beq
\mathcal{A}^{D}_{\;\mbox{\scriptsize{CS}}}(\{p\}_{m+1}) = 
\sum_{i,j,k=1}^{m+1} \mathcal{A}^B(\{\tilde{p}\}_{m}^{(ijk)}) 
\otimes \mathcal{D}^{(ijk)}_{\;\mbox{\scriptsize{CS}}}
(\{\tilde{p}\}_{m}^{(ijk)},\{p\}_{m+1}) \;.
\eeq
In this scheme, each mapping $\{\tilde{p}\}_{m}^{(ijk)}$ is labeled by 
three parton indexes. We note that for large numbers $m$ of external 
partons, the number of mappings and matrix elements required by the 
calculation scales cubically: $N_{\mbox{\scriptsize{CS}}} \sim m^3$. 
For comparison, in the Nagy-Soper (NS) scheme one has
\bea
\mathcal{A}^{D}_{\;\mbox{\scriptsize{NS}}}(\{p\}_{m+1}) & = 
& \sum_{i,j,k=1}^{m+1} \mathcal{A}^B(\{\tilde{p}\}_{m}^{(ij)}) 
\otimes \mathcal{D}^{(ijk)}_{\;\mbox{\scriptsize{NS}}}(\{\tilde{p}\}_{m}^{(ij)},
\{p\}_{m+1})  \\
& = &  \sum_{i,j} \mathcal{A}^B(\{\tilde{p}\}_{m}^{(ij)}) \otimes 
\left( \sum_{k} \mathcal{D}^{(ijk)}_{\;\mbox{\scriptsize{NS}}}
(\{\tilde{p}\}_{m}^{(ij)},\{p\}_{m+1})  \right) 
\label{eq:NSsubtractionterms} \;,
\eea
namely each mapping is characterized by two labels $\{ij\}$ only,
which implies that $N_{\mbox{\scriptsize{NS}}} \sim m^2$. Thus, the
number of mappings and subsequent matrix element re-evaluations is
reduced asymptotically by a factor $m$ compared to CS, which makes the
NS scheme most appealing for processes with large jet
multiplicities. As will be clear in Section \ref{sec:NSformalism}, the
reduced number of subtraction terms is achieved at the price of a more
complicated $\{\tilde{p}\}_{m}$ kinematics. This, in turn, makes the
integration over the phase space of the unresolved parton more
challenging. This sort of complementarity of features provides a
strong motivation for a comparative analysis of the two subtraction
methods, that is one goal of the present paper.

In the Nagy-Soper scheme, the overall structure of the calculation for
the genuine real-emission contribution to $\sigma_{\rm NLO}$ takes the
following form: \bea \sigma_{\rm RE} & \equiv & \int_{m+1} d\Phi_{m+1} \,\,
[ \,\, \mathcal{A}^{R}(\{p\}_{m+1}) \,\, F_{m+1} \nonumber \\ & &  -
  \sum_{i,k,j=1}^{m+1} \mathcal{A}^B(\{\tilde{p}\}_{m}^{(ij)}) \otimes
  \mathcal{D}^{(ijk)}(\{\tilde{p}\}_{m}^{(ij)},\{p\}_{m+1}) \,\, F_{m}
  \,\, ]  \label{eq:subtrreal} \\ & + & \sum_{i,k=1}^{m} \, \int_m
d\Phi_{m} \,\, \mathcal{A}^B(\{p\}_m) \otimes
\mathbf{I}^{(ik)}(\epsilon,\{p\}_{m}) \,\, F_{m}  \label{eq:Iop} \\ &
+ & \sum_{i=\{a,b\}} \, \sum_{k=1}^{m}\, \int_0^1 dx \int_m
d\Phi_{m}(x) \,\, \mathcal{A}^B(x,\{p\}_m) \otimes \left[
  \mathbf{K}^{(ik)}(x,\{p\}_{m}) \right. \nonumber \\  & & +
  \left. \mathbf{P}^{(ik)}(x,\mu_F^2) \right] F_{m} . \label{eq:KPop}
\eea 
There are three kinds of contributions: the first one,
(\ref{eq:subtrreal}) is the \textit{subtracted real} part, while the
remaining two represent the \textit{integrated subtraction
  terms}. Using standard notation, (\ref{eq:Iop}) is the universal
$\mathbf{I}(\epsilon)$ operator, which encodes the full soft/collinear
structure of the matrix element in the form of single and double
$\epsilon$ poles, together with a finite part. The third one,
Eq.(\ref{eq:KPop}), is the $\mathbf{KP}$ operator, which  consists of
purely finite pieces coming from the initial-state splitting
($\mathbf{K}$) as well as from the collinear counterterms
($\mathbf{P}$). It involves an additional integration over the
momentum fraction $x$ of an incoming parton after splitting. The
overall sum $\sigma_{\rm RE}$ of the three contributions is independent on
the subtraction scheme employed.

%-------------%-------------%-------------%-------------
%

%
%-------------%-------------%-------------%-------------

\section{The Nagy-Soper formalism}
\label{sec:NSformalism}

\subsection{Notation}
\label{sec:notation}
First of all, it is necessary to set up a notation for the description
of the parton kinematics and splitting. Let us consider a generic
reaction involving $m+1$ external partons, where we label particles
and the corresponding momenta by an index which takes values $l \in
\{a,b\}$ for the initial state, and $l \in \{ 1, \cdots, m-1 \}$ for
the final state. Let $Q$ be the total momentum: \beq Q = p_a + p_b =
\sum_{l=1}^{m-1} p_l \;.  \eeq In the Nagy-Soper formalism, each
subtraction term highlights two partonic momenta $p_i$ and $p_j$ out
of the $m+1$ available, in such a way that it reproduces the divergent
structure of the amplitude when the two momenta reach the
soft/collinear limits. These two momenta uniquely characterize the
mapping $\{\tilde{p}\}_{m}^{(ij)}$ and thus the subtraction term
itself. We will refer to $p_i,p_j$ as to the \textit{splitting
  momenta} while the remaining partons, also called
\textit{spectators}, will be labeled with $\{k_1, \cdots, k_{m-1}\}$.
As to the splitting momenta, we will always assume $p_j^2=0$, whereas
$p_i$ may have an arbitrary mass $p_i^2 = m_i^2$ \footnote{We point
  out that, in general, also the splitting parton $p_j$ can have a
  mass, $p_j^2 = m_j^2$. In QCD, this is the case of the splitting $g
  \to Q\bar{Q}$, where $Q$ is a massive quark. Note however that in
  this case, also known as \textit{quasi-collinear} splitting, the
  quark mass regulates the collinear divergence, while there is no
  soft singularity. The quasi-collinear splitting functions are thus
  fully regular, although enhanced by logarithms of $m_j^2$ which
  could limit the convergence of the numerical integration in case of
  very small values of $m_j$. With the goal of improving stability,
  one may implement additional local subtraction terms to cope with
  such regular splittings as well, as done for example in
  \cite{Catani:2002hc}. We leave this implementation for future
  developments.}.  Furthermore, we introduce the following quantities
\bea P_{ij} & \equiv & p_i + p_j \;\;\;\;\;\;\;\;   \mbox{for }   i
\in \{ 1, \cdots, m-1 \} \\ P_{ij} & \equiv & p_i - p_j
\;\;\;\;\;\;\;\;   \mbox{for }   i \in \{ a,b \} \;.  \eea \bea K &
\equiv & Q - P_{ij} \;\;\;\;\;\;\;\;      \mbox{for }   i \in \{ 1,
\cdots, m-1 \} \\ K & \equiv & Q - p_j     \;\;\;\;\;\;\;\;\;\,
\mbox{for }   i \in \{ a,b \}  \;.  
\eea 
A divergence occurs when
$\Pijsq \to \misq$, that is when $p_j$ is arbitrarily soft and/or, in
case $\misq=0$, when it is arbitrarily collinear to $p_i$. $K$ is the
so-called \textit{collective spectator}  and coincides with the sum of
all spectator momenta.

\subsection{Momentum mapping}
The momentum mapping $\{p\}_{m+1} \to \{\tilde{p}\}_{m}^{(ij)}$ relates 
momenta before and after the splitting in the following way:
\bea
P_{ij} & \to & \tilde{p}_i \;, \\
k_l & \to & \tilde{k}_l \;\;,\;\;\; l \in \{1,\cdots,m-1\} \;.
\eea
Two basic constraints must be satisfied. First, all partons must be 
on-shell according to the rules
\bea
\tilde{p}_i^2  & = &  p_i^2  \;,   \label{eq:constraint2} \\
\tilde{k}_l^2 & = & k_l^2  \;\;,\;\;\; l \in \{1,\cdots,m-1\} \;.
\eea
For clarity, we remark that the relation $\tilde{p}_i^2  =  p_i^2$ is true 
for all QCD splittings with the exception of $g \to Q\bar{Q}$, where $Q$ is 
a massive quark. As already mentioned, at this stage we do not associate 
any subtraction term to this kind of (non-singular) splittings. For this 
reason we can take (\ref{eq:constraint2}) as a general rule in the context
 of this paper.

As a second constraint, the mapping must reduce to a  simple identity in 
the singular limits, namely
\beq
P_{ij}^2 = \misq \;\;\;\; \Rightarrow \;\;\;\;  \tilde{p}_i = 
P_{ij} \;\;,\;\; \tilde{k}_i = k_i \;.
\eeq
It should be clear that the identity $\tilde{p}_i=P_{ij}$ is no longer 
compatible with the onshellness condition (\ref{eq:constraint2}) away from 
the singular limits, hence $\tilde{p}_i$ requires in general an additional 
contribution in order to remain on-shell. The extra momentum components are 
taken from the spectator partons in a way that is not unique, depending 
whether the splitting occurs in the initial or in the final state.

We begin our exposition considering the case where the splitting parton 
belongs to the final state. 
Following Ref. \cite{Nagy:2007ty}, the momentum of the splitting parton 
is defined to lie on the $Q$-$P_{ij}$ plane according to the formula
\beq
\label{eq:mapping_final}
P_{ij} = \beta \, \tilde{p}_i + \gamma \, Q \;,
\eeq
where the parameters $\beta$ and $\gamma$ are uniquely fixed by setting
\bea
\tilde{Q} & = & Q \;, \\
\tilde{K}^2 & = & K^2  \label{eq:constrain3}
\eea
together with the constraint $\tilde{p}_i^2=\misq$. We remind that all 
kinematical quantities marked with the symbol "\~{}" belong to the mapped 
$m$-parton phase space. In other words, the Nagy-Soper mapping acts in such 
a way that the total momentum $Q$ and the invariant mass $K^2$ of the 
collective spectator are left unchanged. This gives
\bea
\beta & = & 2 \, \sqrt{\frac{(\PijQ)^2-\Pijsq \, \Qsq}{(\misq + 2 \, 
\PijQ-\Pijsq)^2 - 4 \, \misq \, \Qsq}} \;,  \label{eq:beta} \\
& & \nonumber \\
\gamma & = & \frac{2\,\PijQ + \beta \, 
(\Pijsq - 2 \, \PijQ -\misq)}{2 \, \Qsq} \;.
\eea
In the singular limit one gets $\Pijsq = \misq$, $\beta = 1$ and $\gamma = 0$, 
thus $\tilde{p}_i = P_{ij}$ as required. As to the spectator partons, 
Eq.(\ref{eq:constrain3}) suggests that the transformation acting on $K$ 
must have the form of a Lorentz transformation $\Lambda$. In fact, there 
is no room for other possibilities if the mapping is expected to change 
an arbitrary number of spectator momenta at the same time, preserving 
their onshellness. Linear as it is, the same Lorentz transformation 
must also act on each individual spectator:
\bea
K^\mu & = & \Lambda^\mu_{\;\;\nu} \, \tilde{K}^\nu \;, \\
k_i^\mu & = & \Lambda^\mu_{\;\;\nu} \, \tilde{k}_i^{\;\nu} \;\;,\;\;\;\; 
i \in \{1,\cdots,m-1\} \;.
\eea
An explicit representation of $\Lambda$ is given in Ref. \cite{Nagy:2007ty}
\beq \label{eq:Lambda}
\Lambda(K,\tilde{K})^{\mu}_{\;\;\nu} = g^{\mu}_{\;\;\nu} - 
\frac{2 (K + \tilde{K})^{\mu} (K + \tilde{K})_{\nu}}{(K + \tilde{K})^2} + 
\frac{2 K^{\mu} \tilde{K}_{\nu}}{K^2} \;,
\eeq
where 
\bea
K & = & Q - P_{ij} \;, \\
\tilde{K} & = & Q - \tilde{p}_i \;. 
\eea
Note again that, in the singular limit $\Pijsq=\misq$, one gets $\tilde{K}=K$ 
and the mapping reduces to a simple identity.
Incidentally, we observe that (\ref{eq:Lambda}) is only one of many possible 
representations:  indeed, any transformation $\Lambda^\prime = 
\Lambda \cdot R_\alpha(\tilde{K})$,  where $R_\alpha(\tilde{K})$ is a 
spatial rotation of angle $\alpha$ around the direction of $\tilde{K}$, 
is acceptable under the condition that $\alpha \to 0$ in the singular limits.

We now turn to the case where the splitting involves an initial-state parton. 
We take such partons to be on-shell, massless and with zero transverse momentum,
\beq
p_a^2 = p_b^2 = \tilde{p}_a^2 = \tilde{p}_b^2 = 0 \;.
\eeq
Let us assume that parton $a$ splits, 
$\tilde{p}_a \to p_a - p_j \equiv P_{aj}$. The Nagy-Soper mapping leaves 
the other initial-state parton unchanged,
\beq
\tilde{p}_b = p_b \;,
\eeq
while it acts on the final state in a way that preserves the invariant 
mass $K^2$ of the collective spectator. However, unlike the case of 
final-state splitting, the total momentum $Q=p_a+p_b$ is not preserved here. 
The additional constraint necessary to fix uniquely the 
$\{\tilde{p}\}_{m}^{(aj)}$ kinematics is given by the zero transverse 
momentum of the splitting parton, which implies that the  momenta before 
and after the splitting are related by a simple rescaling:
\beq
\label{eq:x}
\tilde{p}_a = x \, p_a \;.
\eeq
Here the variable $x \in [0,1]$ parameterizes the softness of the 
initial-state splitting, $x\to1$ representing the soft limit. Now 
the momenta of the collective spectator before and after the splitting 
are given by
\bea
\tilde{K} & = & \tilde{p_a} + \tilde{p_b} \;, \\
K & = & p_a + p_b - p_j \;,
\eea
so the constraint $\tilde{K}^2=K^2$ allows to fix the value of 
$\tilde{p}_a$ through
\beq
x = \frac{K^2}{Q^2} \;.
\eeq
As to the spectator partons, the Lorentz transformation relating momenta 
before and after the splitting takes again the  form (\ref{eq:Lambda}). 
Note that in the soft and collinear limits, where one can write $p_j$ to 
be proportional to $p_a$, namely $p_j = \kappa\,p_a$, one 
has $K^2 = (1-\kappa) Q^2$, therefore $1-\kappa = x$ and the
 mapping $P_{aj} \equiv p_a - p_j \to \tilde{p}_a$ reduces to an identity.

\subsection{Splitting functions}
\label{sec:splittingfunctions}
We now proceed to derive the splitting functions 
$\mathcal{D}^{(ijk)}(\{\tilde{p}\}_{m}^{ij},\{p\}_{m+1})$ in the Nagy-Soper 
formalism. It is convenient to take spin correlations in evidence and 
rewrite Eq.(\ref{eq:NSsubtractionterms}) in the form
\bea
\mathcal{A}^{D} & = & \sum_{i,j,k=1}^{m+1} 
\mathcal{A}^B(\{\tilde{p}\}_{m}^{(ij)}) \otimes  
\mathcal{D}^{(ijk)}(\{\tilde{p}\}_{m}^{(ij)},\{p\}_{m+1})  \\
& = &  \sum_{i,j,k=1}^{m+1} \sum_{\tilde{s}_1,\tilde{s}_2=\pm} 
\mathcal{A}^B_{\tilde{s}_1 \tilde{s}_2}(\{\tilde{p}\}_{m}^{(ij)}) \; 
\mathcal{D}^{(ijk)}_{\tilde{s}_1 \tilde{s}_2}(\{\tilde{p}\}_{m}^{(ij)},\{p\}_{m+1}) 
\,\, ( \mathbf{T}_i \cdot \mathbf{T}_k ) \;,  \label{eq:subtrterm}
\eea
where $\tilde{s}_1,\tilde{s}_2$ are the helicity values of the splitting 
partons $\tilde{p}_i,\tilde{p}_k$, while $(\mathbf{T}_i \cdot \mathbf{T}_k)$ 
denotes the color correlator. Note that when $i=k$, the color correlator 
reduces to a simple overall color factor. The function 
$\mathcal{D}^{(ijk)}_{\tilde{s}_1 \tilde{s}_2}$ splits into two kind of 
contributions $W^{(ii,j)}$ and $W^{(ik,j)}$, called 
respectively \textit{diagonal} and \textit{interference} terms:
\beq
\mathcal{D}^{(ijk)}_{\tilde{s}_1 \tilde{s}_2} = 
W^{(ii,j)}_{\tilde{s}_1 \tilde{s}_2}\,\delta_{ik} + 
W^{(ik,j)}_{\tilde{s}_1 \tilde{s}_2} (1-\delta_{ik}) \, 
\delta_{{\tilde{s}_1 \tilde{s}_2}}  \;.  \label{eq:diaginterfterms}
\eeq
At NLO, while the diagonal terms $W^{(ii,j)}_{\tilde{s}_1 \tilde{s}_2}$ show both 
soft and collinear singularities, the interference terms 
$W^{(ik,j)}_{\tilde{s}_1 \tilde{s}_2}$ can only be divergent in the soft 
limit, since the jet function excludes any configuration where partons 
$p_i,p_k$ are simultaneously collinear to $p_j$. The factor 
$\delta_{\tilde{s}_1 \tilde{s}_2}$ in the interference term is justified 
by the fact that, in the soft limit, the two  partons connected by a 
soft gluon exchange have the same spin index.

Let us consider a generic QCD matrix element $\mathcal{M}_{m+1}$ 
characterized by $m+1$ external legs, and look at the way it factorizes 
when an external parton splits. We take the case of a final-state 
splitting as an example. Depending on flavor, there are three possible 
splittings: $q \to qg$, $g \to q\bar{q}$ and $g \to gg$. The corresponding 
matrix elements read:
\bea
\label{eq:Mm+1_qqg}
\mathcal{M}_{m+1}^{q \to qg} & = & \mathcal{H}_{m} \, 
\frac{\s{P}_{ij}+m_i}{\Pijsq - \misq} \, \left( \sqrt{4\pi\as} \, 
\mathbf{T}_i \, \right) \s{\varepsilon}^*(p_j,s_j) \, u(p_i,s_i) \;, \\
\label{eq:Mm+1_gqq}
\mathcal{M}_{m+1}^{g \to q\bar{q}} & = & \mathcal{H}_{m}^{\mu} \, 
\frac{D_{\mu\nu}(P_{ij},n_i)}{\Pijsq} \,\, \bar{u}(p_j,s_j) 
\left( \sqrt{4\pi\as} \, \mathbf{T}_i \, \right) \gamma^{\nu} \, 
u(p_i,s_i) \;, \\
\label{eq:Mm+1_ggg}
\mathcal{M}_{m+1}^{g \to gg} & = & \mathcal{H}_{m}^{\mu} \, 
\frac{D_{\mu\nu}(P_{ij},n_i)}{\Pijsq} \left( \sqrt{4\pi\as} \, 
\mathbf{T}_i \, \right) G_{\nu\rho\sigma} \, \varepsilon^{\rho}(p_i,s_i)^* \, 
\varepsilon^{\sigma}(p_j,s_j)^* \;.
\eea
We generically indicate with $\mathbf{T}_i$ the color matrix associated 
with the splitting $\tilde{p}_i \to p_i+p_j$. Furthermore, $\varepsilon(p,s)$ 
and $u(p,s)$ represent polarization vectors and spinors for a given 
assignment of momentum $p$ and helicity $s$. $\mathcal{H}_{m}$ is a 
$m$-parton subamplitude, and $G^{\nu\rho\sigma}$ denotes the triple gluon 
vertex
\beq
G^{\nu\rho\sigma} \equiv g^{\nu\rho}(p_j-p_i)^\sigma + 
g^{\rho\sigma}(p_i+P_{ij})^\nu + g^{\sigma\nu}(-P_{ij}-p_j)^\rho  \;.
\eeq
Finally, $D^{\mu\nu}$ is the numerator of the gluonic propagator in the 
axial gauge
\beq
D^{\mu\nu}(P,n) = -g^{\mu\nu} + \frac{P^{\mu} n_{i}^{\nu} + P^{\nu} n_{i}^{\mu}}{P 
\cdot n_i} \;,
\eeq
where $n_i$ is an arbitrary light-like vector whose definition closely 
follows Ref. \cite{Nagy:2007ty} \footnote{In the original formulation by 
Nagy and Soper \cite{Nagy:2007ty}, the initial-state partonic   
momenta $p_a,p_b$ appearing in Eq.(\ref{eq:nidef}) are replaced by the 
corresponding \textit{hadronic} momenta, denoted $p_A,p_B$. We note that 
either choice is equivalent for the purposes of subtraction, since the 
vector $n_i$ always appears in ratios, and any possible dependence upon 
the hadronic momentum fractions fully cancels in the end.}
\beq
\label{eq:nidef}
n_i \equiv 
\begin{cases}
p_b \;\;,&  \mbox{for} \;\; i = a\;\;, \\
p_a \;\;,&  \mbox{for} \;\; i = b\;\;, \\
\displaystyle{
Q
-\frac{Q^2}
{Q\!\cdot\! \tilde{p}_i
+ \sqrt{(Q\!\cdot\! \tilde{p}_i)^2 - Q^2\, \misq}}\,\tilde{p}_i
}
\;\;,&  \mbox{for} \;\; i \in \{1,\dots,m-1\}\;\;.
\end{cases}
\eeq

The subsequent step is to find out expressions which approximate the 
divergent behavior of Eqs.(\ref{eq:Mm+1_qqg}), (\ref{eq:Mm+1_gqq}), 
(\ref{eq:Mm+1_ggg}). One can start by inserting simple identities on 
the right side of the subamplitude $\mathcal{H}_m$:
\bea
1& = & \frac{(\s{\tilde{p}}_i+m_i)\,\s{n}_i + \s{n}_i 
(\s{\tilde{p}}_i-m_i)}{2 \, \tilde{p}_i \cdot n_i} \;, \label{eq:id1} \\
g^{\alpha\mu} & = & -D^{\alpha\mu}(\tilde{p}_i,n_i) + 
\frac{\tilde{p}_i^{\alpha}n_i^{\mu} + 
n_i^{\alpha}\tilde{p}_i^{\mu}}{\tilde{p}_i \cdot n_i}  \label{eq:id2}  \;.
\eea
The first identity is used for the $q \to qg$ splitting, while the second 
one applies to the other two cases, $g \to q\bar{q}$ and $g \to gg$. We 
remind that $\tilde{p}_i$ is an on-shell vector with 
$\tilde{p}_i^2=p_i^2=\misq$. In Ref. \cite{Nagy:2007ty}, it has been 
proved explicitly that the contribution coming from the second term 
in both identities does not lead to any collinear or soft singularity 
and can thus be discarded. There is a simple way of understanding this 
property, based on equations of motion and Ward identity. Indeed, in 
the singular limit $\Pijsq \to \misq$, where $P_{ij} \approx \tilde{p}_i$, 
one can write
\bea
\s{P}_{ij}+m_i & \approx & \sum_{\tilde{s}_i=\pm} u(\tilde{p}_i,\tilde{s}_i) \, 
\overline{u}(\tilde{p}_i,\tilde{s}_i) \;, \\
D^{\alpha\mu}(P_{ij},n_i) & \approx & \sum_{\tilde{s}_i=\pm} 
\varepsilon^{\alpha}(\tilde{p}_i,\tilde{s}_i)^* \varepsilon^{\mu}
(\tilde{p}_i,\tilde{s}_i) \;.
\eea
It immediately follows that the second term in (\ref{eq:id1}) gives a 
vanishing contribution due to the equation of motion $(\s{\tilde{p}_i}-m_i) \, 
u(\tilde{p}_i,\tilde{s}_i) = 0$. Similarly, the second term in (\ref{eq:id2}) 
vanishes thanks to Ward identity, $\tilde{p}_i^\mu \mathcal{H}_\mu = 0$. Note 
that $n_i^\mu \mathcal{H}_\mu = 0$ in the axial gauge we are considering.

Neglecting the second term from the identities, one is left with the following 
expressions which approximate the divergent behavior of the original matrix 
elements:
\bea
\label{eq:Mn+1_qqg_sing}
\mathcal{M}_{m+1}^{q \to qg,\,div} & = & \mathcal{H}_{m} \, 
\frac{(\s{\tilde{p}}_i+m_i)\,\s{n}_i}{2 \, \tilde{p}_i \cdot n_i} \, 
\frac{\s{P}_{ij}+m_i}{\Pijsq - \misq} \, \left( \sqrt{4\pi\as} \, 
\mathbf{T}_i \, \right) \s{\varepsilon}^*(p_j,s_j) \, u(p_i,s_i) \,, \\
\label{eq:Mm+1_gqq_sing}
\mathcal{M}_{m+1}^{g \to q\bar{q},\,div} & = & -\mathcal{H}_{m}^{\alpha} \, 
D_{\alpha\mu}(\tilde{p}_i,n_i) \, \frac{D^{\mu\nu}(P_{ij},n_i)}{\Pijsq} \, 
\bar{u}(p_j,s_j) \, \left( \sqrt{4\pi\as} \, \mathbf{T}_i \, \right) 
\gamma^{\nu} \, u(p_i,s_i)  \,, \\
\label{eq:Mm+1_ggg_sing}
\mathcal{M}_{m+1}^{g \to gg,\,div} & = & \mathcal{H}_{m}^{\alpha} \, 
D_{\alpha\mu}(\tilde{p}_i,n_i) \, \frac{D^{\mu\nu}(P_{ij},n_i)}{\Pijsq} 
\left( \sqrt{4\pi\as} \, \mathbf{T}_i \, \right) G_{\nu\rho\sigma} 
\,\varepsilon^{\rho}(p_i,s_i)^* \varepsilon^{\sigma}(p_j,s_j)^* .
\eea
Now using
\bea
\s{\tilde{p}}_i+m_i & = & \sum_{\tilde{s}_i=\pm} u(\tilde{p}_i,\tilde{s}_i) \, 
\overline{u}(\tilde{p}_i,\tilde{s}_i) \;, \label{eq:uu} \\
D^{\alpha\mu}(\tilde{p}_i,n_i) & = & \sum_{\tilde{s}_i=\pm} 
\varepsilon^{\alpha}(\tilde{p}_i,\tilde{s}_i)^* \varepsilon^{\mu}(\tilde{p}_i,
\tilde{s}_i) \label{eq:epseps} \;,
\eea
and associating the first factor in (\ref{eq:uu}), (\ref{eq:epseps}) with the 
subamplitude $\mathcal{H}_m$, one gets a product of a complete $m$-parton 
scattering amplitude times a genuine splitting function, \textit{i.e.} 
$\mathcal{M}_{m+1}^{div} = \mathcal{H}_m \otimes  v$. Once the contribution 
of color degrees of freedom is factored out, the splitting functions take 
the final form:
\bea
v^{(ij)}_{\tilde{s}_i s_i s_j}(q \to qg) & = & \frac{\sqrt{4\pi\as}}{\Pijsq - 
\misq} \, \frac{\bar{u}(\tilde{p}_i,\tilde{s}_i) \, \s{n}_i (\s{P}_{ij} +
 m_i) \, \s{\varepsilon}(p_j,s_j) \, u(p_i,s_i)}{2 \, \tilde{p}_i \cdot n_i}  
\;, \label{eq:splf_qqg} \\
v^{(ij)}_{\tilde{s}_i s_i s_j}(g \to q\bar{q}) & = & - 
\frac{\sqrt{4\pi\as}}{\Pijsq} \, \varepsilon^{\mu}(\tilde{p}_i,\tilde{s}_i) 
D_{\mu\nu}(P_{ij},n_i) \, \bar{u}(p_j,s_j) \, \gamma^\nu \, u(p_i,s_i) \;, 
\label{eq:splf_gqq} \\
v^{(ij)}_{\tilde{s}_i s_i s_j}(g \to gg) & = & \frac{\sqrt{4\pi\as}}{\Pijsq} 
\, D^{\mu\nu}(P_{ij},n_i) \, G_{\nu\rho\sigma} \,  
\varepsilon_{\mu}(\tilde{p}_i,\tilde{s}_i) \, \varepsilon^{\rho}(p_i,s_i)^* \, 
\varepsilon^{\sigma}(p_j,s_j)^* \;.  \label{eq:splf_ggg}
\eea
Note that functions (\ref{eq:splf_qqg}) and (\ref{eq:splf_ggg}) show both  
collinear and soft divergences, whereas in the case of $g \to q\bar{q}$ 
splitting, Eq.(\ref{eq:splf_gqq}), the divergence can only be collinear. 
In all cases where the splitting functions reach the soft limit, gluon 
emission can be treated in the eikonal approximation, and the splitting 
function reduces to a much simpler expression:
\beq
\label{eq:eikonal}
v^{(ij),\,eik}_{\tilde{s}_i s_i s_j} = \sqrt{4\pi\as} \, \delta_{\tilde{s}_i,s_i} 
\frac{\varepsilon(p_j,s_j)^* \cdot p_i}{p_i \cdot p_j} \;.
\eeq
The derivation of the splitting functions in the case of initial-state 
radiation is quite similar, the only formal difference being in the form 
of the spinors and polarization vectors, and possibly in an overall sign. 
A comprehensive list of splitting functions, including the case of 
initial-state emission, is reported in  Ref. \cite{Nagy:2007ty}. 

In the last step, the splitting functions are combined to give the diagonal 
and interference  contributions to the dipole 
$\mathcal{D}^{(ijk)}_{\tilde{s}_1\tilde{s}_2}$ as to Eq.(\ref{eq:diaginterfterms}):
\bea
W^{(ii,j)}_{\tilde{s}_1\tilde{s}_2} & = & \sum_{s_i,s_j} v^{(ij)}_{\tilde{s}_1 s_i s_j} 
\left( v^{(ij)}_{\tilde{s}_2 s_i s_j} \right)^*  \label{eq:diagterm} \\
W^{(ik,j)}_{\tilde{s}_1\tilde{s}_2} & = & 
\sum_{s_i,s_j} v^{(ij),\,eik}_{\tilde{s}_1 s_i s_j} 
\left( v^{(kj),\,eik}_{\tilde{s}_2 s_i s_j} \right)^* \,
\delta_{\tilde{s}_1\tilde{s}_2} \;.  \label{eq:interfterm}
\eea
Note that there is an ambiguity with the definition of the 
interference term $W^{(ik,j)}$, in that one can adopt either the 
mapping $\{\tilde{p}\}_{m}^{(ij)}$ or $\{\tilde{p}\}_{m}^{(kj)}$ to 
compute it. Instead of choosing one, it is also possible to take a 
weighted combination of the two:
\beq
W^{(ik,j)} = A_{ik} \, W^{(ik,j)}_{[i]} + A_{ki} \, W^{(ik,j)}_{[k]}  \;.
\eeq
Here the subscript $[i] \, ([k])$ means that the corresponding term is 
evaluated considering the mapping $\{\tilde{p}\}_{m}^{(ij)} \, 
(\{\tilde{p}\}_{m}^{(kj)})$. The weight factors always satisfy the condition
\beq
A_{ik} + A_{ki} = 1 \;.
\eeq
One motivation for taking such a combination, as discussed in detail 
in \cite{Nagy:2008ns,Nagy:2008eq}, is that it shows more favorable 
properties in the context of the parton shower. 
Including the color factor, the net contribution summed over the two 
graphs arising from the interference of soft gluons emitted from partons 
$i$ and $k$ is
\bea
& & W^{(ik,j)} \, \left( \mathbf{T}_i \cdot \mathbf{T}_k \right) + W^{(ki,j)} 
\, \left( \mathbf{T}_k \cdot \mathbf{T}_i  \right) \\
& = & A_{ik} \left[ W^{(ik,i)}_{[i]} \, \left( \mathbf{T}_i \cdot 
\mathbf{T}_k \right)  + W^{(ki,j)}_{[i]} \, \left( \mathbf{T}_k \cdot 
\mathbf{T}_i \right) \right] \\
& + & A_{ki} \left[ W^{(ik,i)}_{[k]} \, \left( \mathbf{T}_i \cdot 
\mathbf{T}_k \right) + W^{(ki,j)}_{[k]} \,  \left( \mathbf{T}_k \cdot 
\mathbf{T}_i \right) \right]  \;.
\eea
Note that the weight factors can be either constants or functions of 
$\{p\}_{m+1}$. Several expressions for them have been proposed in the 
literature. In our implementation, we consider the definition given by 
Eq.(7.12) of Ref. \cite{Nagy:2008eq}:
\beq
A_{ik} = \frac{(p_j \cdot p_k) 
(p_i \cdot Q)}{(p_j \cdot p_k) (p_i \cdot Q) + (p_i \cdot p_j) (p_k \cdot Q)} \;.
\eeq

%-------------%-------------%-------------%-------------
%

\subsection{Phase space factorization}
\begin{figure}
\begin{center}
\begin{flushleft}
\underline{Nagy-Soper}
\end{flushleft}
\includegraphics[width=0.23\textwidth]{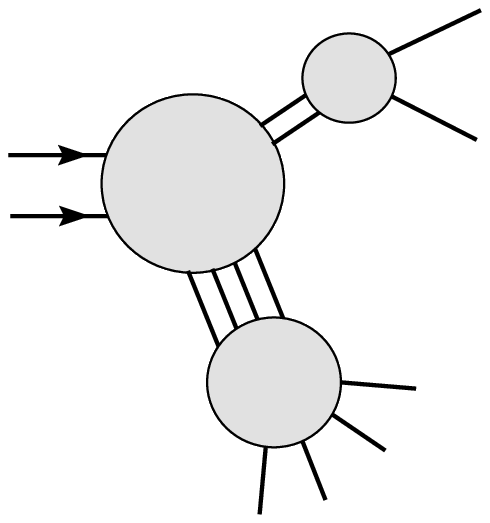} 
\hspace{0.18\textwidth} \includegraphics[width=0.19\textwidth]{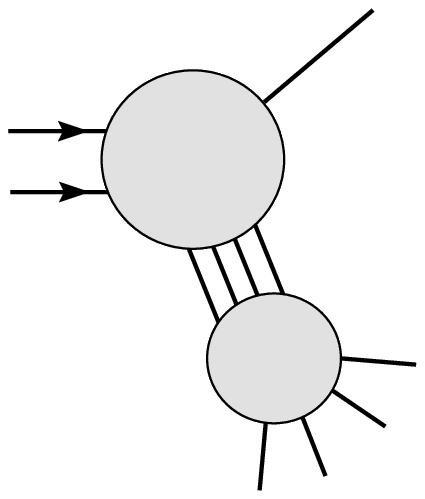} 
\put(3,85){\includegraphics[width=0.1\textwidth]{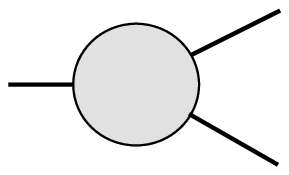}}
\put(-7,95){\large{$\otimes$}}
\put(-231,65){\large{$Q$}}
\put(-201,88){{$P_{ij}$}}
\put(-164,104){\large{$p_i$}}
\put(-164,75){\large{$p_j$}}
\put(-214,23){\large{$K$}}
\put(-277,74){\large{$p_a$}}
\put(-277,58){\large{$p_b$}}
\put(-28,95){\large{$\tilde{p}_i$}}
\put(48,109){\large{$p_i$}}
\put(48,81){\large{$p_j$}}
\put(-49,62){\large{$Q$}}
\put(-33,22){\large{$\tilde{K}$}}
\put(-95,72){\large{$p_a$}}
\put(-95,56){\large{$p_b$}}
\put(-136,51){\Large{$\Rightarrow$}} \\
\vspace{0.5cm}
$\{p_i,p_j\} \to \tilde{p}_i$ \;;\;
$\{K,Q\} \to \{\tilde{K},Q\}$ \\
$p_i+p_j+K = \tilde{p}_i+\tilde{K}$
\begin{flushleft}
\underline{Catani-Seymour}
\end{flushleft}
\includegraphics[width=0.28\textwidth]{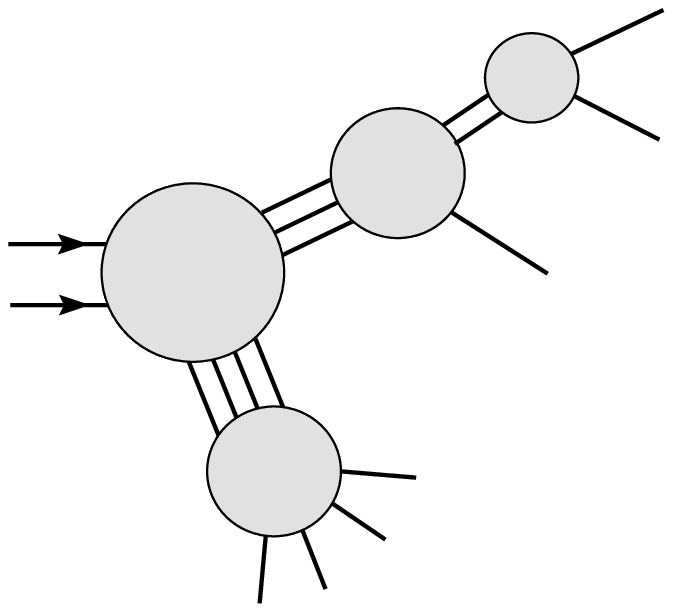} \hspace{0.12\textwidth} 
\includegraphics[width=0.22\textwidth]{phsp_2.eps} 
\put(12,89){\includegraphics[width=0.1\textwidth]{phsp_3.eps}}
\put(2,98){\large{$\otimes$}}
\put(-15,103){\large{$\tilde{p}_i$}}
\put(57,113){\large{$p_i$}}
\put(57,85){\large{$p_j$}}
\put(-15,65){\large{$\tilde{p}_k$}}
\put(-62,61){\large{$Q$}}
\put(-46,22){\large{$R$}}
\put(-107,71){\large{$p_a$}}
\put(-107,57){\large{$p_b$}}
\put(-150,109){\large{$p_i$}}
\put(-150,82){\large{$p_j$}}
\put(-185,93){{$P_{ij}$}}
\put(-172,56){\large{$p_k$}}
\put(-212,76){\large{$P_{ijk}$}}
\put(-244,56){\large{$Q$}}
\put(-230,19){\large{$R$}}
\put(-287,66){\large{$p_a$}}
\put(-287,52){\large{$p_b$}}
\put(-140,46){\Large{$\Rightarrow$}} \\
\vspace{0.5cm}
$\{p_i,p_j\} \to \tilde{p}_i$ \;;\;
$\{p_k,R,Q\} \to \{\tilde{p}_k,R,Q\}$ \\
$p_i+p_j+p_k = \tilde{p}_i+\tilde{p}_k$
\end{center}
\caption{Schematic representation of the phase space factorization and 
parameterization in terms of recursive splittings: comparison between the 
Nagy-Soper scheme (top) and the Catani-Seymour scheme (bottom). $p_i$ and 
$p_j$ are the splitting momenta, whereas the the remaining momenta are 
classified as spectators. The shown example refers to the case of a 
final-state splitting.}
\label{fig:phsp}
\end{figure}

An important feature of the Nagy-Soper mapping is that all spectator momenta 
are changed at the same time. This fact obviously has an impact on the way 
the factorization of the phase space is achieved. While in the case of 
initial-state emission there are no substantial differences with respect 
to the Catani-Seymour scheme, the factorization is derived in a slightly 
different way when the splitting occurs in the final state.
Let us start with the usual definition of the Lorentz invariant phase space 
for a generic final state with $m+1$ partons, where we closely follow the 
notation introduced in Section \ref{sec:notation}:
\bea
&& d\Phi_{m+1}(p_i,p_j,k_1,\cdots,k_{m-1};Q) \equiv \\ \nonumber 
&& \frac{d^3p_i}{(2\pi)^d\,2p_i^0} \,  \frac{d^3p_j}{(2\pi)^d\,2p_j^0} \,  
\frac{d^3k_1}{(2\pi)^d\,2k_1^0}  \cdots  \frac{d^3k_{m-1}}{(2\pi)^d\,2k_{m-1}^0} 
\, (2\pi)^d \, \delta^d(Q-p_i-p_j-k_1-\cdots-k_{m-1}) \;.
\eea
We re-organize the phase space in terms of recursive splittings, as 
schematically represented in Fig.\ref{fig:phsp}:
\bea
\label{eq:recursivesplitting}
&& d\Phi_{m+1}(p_i,p_j,k_1,\cdots,k_m;Q) = \\ \nonumber
&& \int_{K^2_{min}}^{K^2_{max}} \frac{dK^2}{2\pi} \,  \int_{P^2_{min}}^{P^2_{max}} 
\frac{dP^2_{ij}}{2\pi} \, d\Phi_{m-1}(k_1,\cdots,k_{m-1};K) \, 
d\Phi_2(P_{ij},K;Q) \, d\Phi_2(p_i,p_j;P_{ij})  \;,
\eea
where
\bea
K^2_{min} & = & (m_{k_1}+\cdots+m_{k_{m-1}})^2  \\
K^2_{max} & = & (\sqrt{Q^2}-m_{p_i}-m_{p_j})^2 \\
P^2_{min} & = & (m_{p_i}+m_{p_j})^2 \\
P^2_{max} & = & (\sqrt{Q^2}-\sqrt{K^2})^2 \;.  \label{eq:Pijsq_max}
\eea
Here $m_{p_i},m_{p_j},m_{k_i}$ represent the masses of the on-shell final-state 
partons, while  $\sqrt{K^2}$ is the invariant mass of the collective spectator. 
Looking at the first factor in the integral (\ref{eq:recursivesplitting}), 
one observes that
\beq
 d\Phi_{m-1}(k_1,\cdots,k_{m-1};K) = d\Phi_{m-1}(\tilde{k}_1,\cdots,
\tilde{k}_{m-1};\tilde{K}) \;.
\eeq
This immediately follows from the fact that the mapping $K \to \tilde{K}$ 
is a Lorentz transformation, and the phase space is Lorentz invariant.
 The Jacobian of the second factor is not as trivial, but admits a  simple 
derivation in the frame where the total momentum $Q$ is at rest. In this 
frame, the two-body phase space can be parameterized in terms of angular 
variables,
\beq
d\Phi_2(P_{ij},K;Q) = \frac{1}{8\,(2\pi)^{d-2}} \, 
\frac{\lambda(Q^2,P_{ij}^2,K^2)^{\frac{d-3}{2}}}{(Q^2)^{\frac{d-2}{2}}} \, 
\int d\Omega_{d-1} \;,
\eeq
where $d\Omega_{d-1}$ represents the solid angle in $d$ dimensions and 
$\lambda$ is the standard K\"allen function,
\beq
\lambda(x,y,z) = x^2 + y^2 + z^2 -2xy - 2xz - 2yz \;.
\eeq
It should be clear from Eq.(\ref{eq:mapping_final}) that, when the total 
momentum $Q$ is at rest, the  vector $\vec{\tilde{p}}_i$ is just proportional 
to $\vec{P}_{ij}$. Thus, the two spatial vectors are described by the same 
angular variables, and the integral $\int d\Omega_{d-1}$ for the two phase 
spaces $d\Phi_2(P_{ij},K;Q)$ and $d\Phi_2(\tilde{p}_{i},\tilde{K};Q)$ is the 
same. This implies that the Jacobian related to the mapping $P_{ij} \to 
\tilde{p}_i$ is simply
\beq
d\Phi_2(P_{ij},K;Q) = \left( \frac{\lambda(Q^2,P^2_{ij},K^2)}
{\lambda(Q^2,\misq,K^2)} \right)^{\frac{d-3}{2}} \, 
d\Phi_2(\tilde{p}_i,\tilde{K};Q) \;.
\eeq
In the end, the phase space for the final-state emission can be expressed 
in the fully factorized form 
\beq
d\Phi_{m+1}(p_i,p_j,k_1,\cdots,k_{m-1};Q) = 
d\Phi_m(\tilde{p}_i,\tilde{k}_1,\cdots,\tilde{k}_{m-1};Q) \times d\xi_{fin} \;,
\eeq
where
\beq
\label{eq:dxi_fin}
d\xi_{fin} = \frac{dP^2_{ij}}{2\pi} \, 
\left( \frac{\lambda(Q^2,P^2_{ij},K^2)}{\lambda(Q^2,\misq,K^2)}
\right)^{\frac{d-3}{2}} d\Phi_2(p_i,p_j;
\underbrace{\beta\,\tilde{p}_i+\gamma\,Q}_{P_{ij}})
\eeq
is the measure of the splitting phase space in $d$ dimensions.

As already mentioned, in the case of initial-state emission, the factorization 
of the phase space does not show conceptual differences with Catani-Seymour 
subtraction, where the splitting phase space is written as \cite{Catani:1996vz}
\beq
\label{eq:dxi_ini}
d\xi_{ini} = \frac{d^{d-1}p_j}{(2\pi)^{d-1}\,2p^0_j}\,\Theta(x)\Theta(1-x)\,
\delta(x-\bar{x}(p_j)) \;.
\eeq
The variable $x$ appearing in the formula is the same defined in 
Eq.(\ref{eq:x}), $\tilde{p}_a=x p_a$, namely the one that parameterizes 
the softness of the initial-state radiation. The factor 
$\delta(x-\bar{x}(p_j))$ comes from the fact that the value of $x$ 
is dictated by the kinematics of the unconstrained parton $p_j$ and 
can thus be expressed as a function of it, $\bar{x}(p_j)$.
The explicit form of the integration measures, 
Eqs.(\ref{eq:dxi_fin})-(\ref{eq:dxi_ini}), obviously depends 
on the specific parameterization of the phase space and will be 
clear in Section \ref{sec:splitting_variables}, where we define 
the splitting variables used in our calculation.

%-------------%-------------%-------------%------------- 
%

\section{Integration over the unresolved phase space}

\subsection{Splitting variables}
\label{sec:splitting_variables}
In $d=4$ dimensions, the phase space of the splitting has three degrees 
of freedom. 
One possible way of parameterizing them relies on scalar products and 
splitting variables defined in a Lorentz-covariant fashion, as proposed 
in Ref. \cite{Chung:2010fx}. In fact, this choice is conceptually simple 
and leads to quite compact formulae of the integrated dipoles for massless 
partons \cite{Chung:2012rq}. On the other hand, when one turns to the fully 
massive case, the kinematical bounds of the splitting get more complicated 
and many additional terms are generated in the formulae. With the goal of 
keeping the final expressions reasonably compact, which we find desirable 
for our  semi-numerical approach, we have adopted an alternative 
parameterization that we describe in this Section.

Let us start with the case of a final-state emission. Given a set of 
momenta $\{\tilde{p}\}_m$ as input, we want to construct the splitting 
$\tilde{p}_i \to p_i+p_j$ and thus the set of momenta $\{p\}_{m+1}$ out 
of three parameters that we call \textit{collinear}, \textit{soft} and 
\textit{azimuthal} variables. In the singular limit $P_{ij}^2 \to \misq$, 
the collinear and the soft variables must resemble the two quantities that 
naturally parameterize the collinearity and softness of the splitting, 
\textit{i.e.} the relative angle between the nearly-collinear partons, 
$\theta_{ij}$, and  the energy of the unresolved  parton, $E_j$. By azimuthal 
variable, we mean the second angular parameter that is necessary in order to 
uniquely fix the  kinematics of the splitting.

\begin{figure}
\begin{center}
\includegraphics[width=0.5\textwidth]{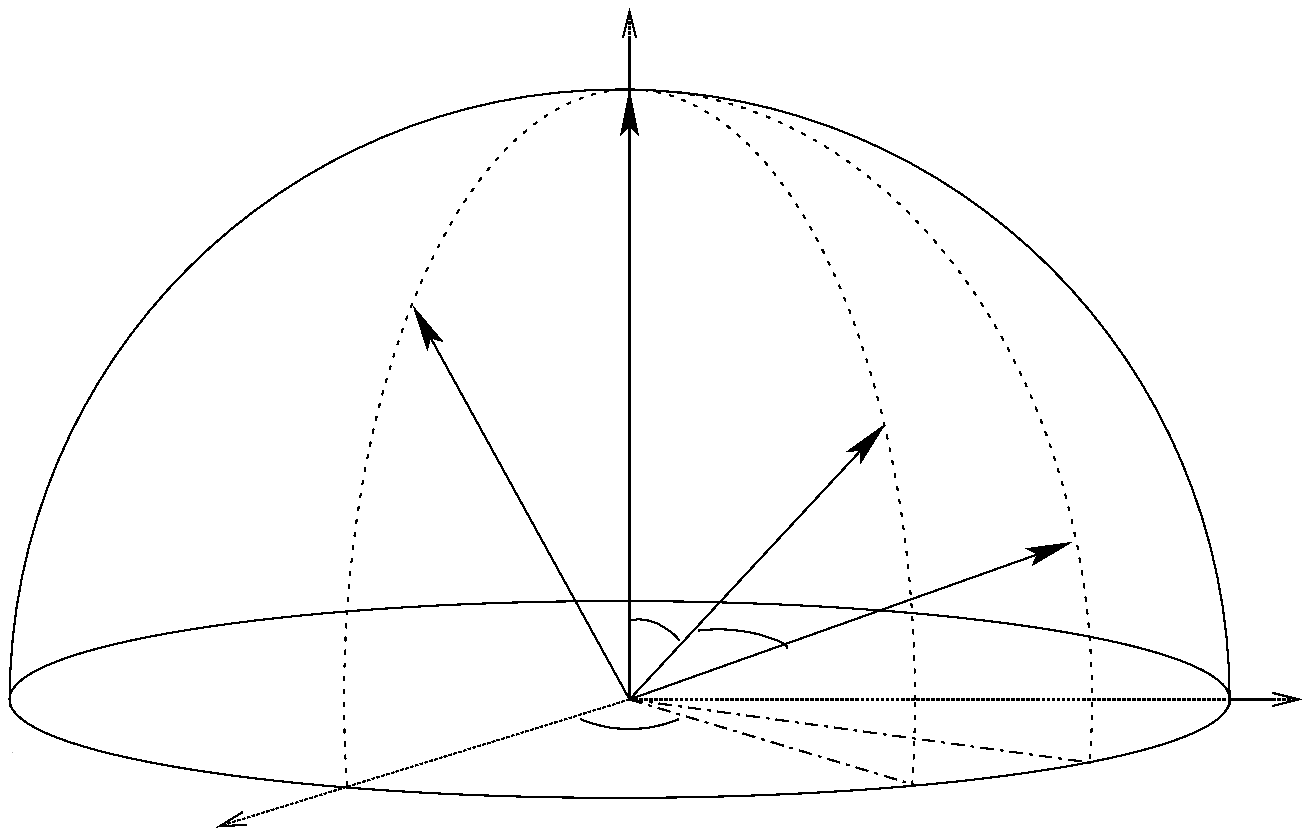}
\put(-31,41){$p_i$}
\put(-60,65){$p_j$}
\put(-103,115){$\tilde{p}_i$}
\put(-154,78){$\tilde{p}_k$}
\put(-105,140){$z$}
\put(-188,-6){$x$}
\put(11,20){$y$}
\put(-111,9){$\phi_j$}
\put(-104,39){$\theta_j$}
\put(-87,37){$\theta_{ij}$}
\end{center}
\caption{Parameterization of the angular variables for the final-state 
splitting $\tilde{p}_i \to p_i+p_j$. Here $\tilde{p}_k$ is the spectator 
parton selected to define the azimuthal variable $\phi_j$.}
\label{fig:angles_fin}
\end{figure}

The definition of the angular variables requires a reference frame to be set. 
We consider an orthogonal set of axes $(x,y,z)$, and identify the 
$z$-axis with the spatial direction of the vector $\tilde{p}_i$. There 
is full freedom in selecting the direction of the $x$-axis: when this is 
done, the azimuthal variable is uniquely determined as the angle $\phi_j$ 
which separates the unresolved parton $p_j$ from the the $x$-$z$ plane. A 
convenient choice is to consider the spectator momentum $\tilde{p}_k$ which 
appears in each interference term as a reference direction, so that the 
$x$-axis lies on the plane $\tilde{p}_i$-$\tilde{p}_k$ as shown in 
Figure \ref{fig:angles_fin}. This choice obviously helps to keep the 
dependence of the integrands upon the azimuthal variable as simple as 
possible. In this frame, the angle $\theta_j$ between partons $p_j$ and 
$\tilde{p}_i$ is a good candidate for the collinear variable. Note that 
when $P_{ij}^2 \approx \misq$, one gets indeed $\tilde{p}_i \approx p_i+p_j$ 
and thus $\cos\theta_j \approx \cos\theta_{ij}$.
The definition of the soft variable is more complicated, since the relation 
between the energy $E_j$ and the invariant mass $P_{ij}^2$ is not generally 
invertible. In the Lorentz frame where the total momentum $Q$ is at rest, 
it reads
\beq
\label{eq:Ej_vs_Pijsq}
E_j = \frac{\sqrt{\Qsq}\,(\Pijsq-\misq)}{\Pijsq - \misq + 
2\,\sqrt{\Qsq}\,(\tilde{p}_i^0-\cos\theta_j\,\beta\,\vert 
\vec{\tilde{p}}_i \vert)} \;.
\eeq
The parameter $\beta$ is given by Eq.(\ref{eq:beta}) and is a non-monotonic 
function of $\Pijsq$. In consequence, by choosing $E_j$ as the soft variable, 
one would be led to some ambiguity in the assignment of $\Pijsq$. We observe, 
however, that such ambiguity is  solved if one sets $\beta=1$ 
in (\ref{eq:Ej_vs_Pijsq}). In fact, this defines a new variable $\bar{E}_j$:
\beq
\bar{E}_j \equiv \frac{\sqrt{\Qsq}\,(\Pijsq-\misq)}{\Pijsq - \misq + 
2\,\sqrt{\Qsq}\,(\tilde{p}_i^0-\cos\theta_j\,\vert \vec{\tilde{p}}_i \vert)} \;.
\eeq
Clearly when $P_{ij}^2 \approx \misq$ one has $\beta \approx 1$ and therefore 
$\bar{E}_j \approx E_j$, hence the variable $\bar{E}_j$ is a good candidate 
to be a soft variable. It is further  convenient  to divide it by its 
kinematically allowed maximum value and get a normalized soft variable $e$:
\beq
e \equiv \bar{E}_j/\bar{E}_{j}^{max} \;,
\eeq
where
\beq
\bar{E}_{j}^{max} = \frac{\sqrt{\Qsq}\,(P_{ij,max}^2-\misq)}{P_{ij,max}^2 - 
\misq + 2\,\sqrt{\Qsq}\,(\tilde{p}_i^0-\cos\theta_j\,\vert \vec{\tilde{p}}_i 
\vert)} \;,
\eeq
and $P_{ij,max}^2$ is given by Eq.(\ref{eq:Pijsq_max}).
To summarize, in case of final-state emission,  the integration of the 
splitting phase space runs over the variables 
\bea
e & \in & [0,1] \;, \\
c \equiv \cos\theta_j & \in & [-1,1]  \\
\phi \equiv \phi_j & \in & [0,2\pi] \;. 
\eea
The soft and collinear limits correspond to $e\to0$ and $c\to1$ respectively.

\begin{figure}
\begin{center}
\includegraphics[width=0.5\textwidth]{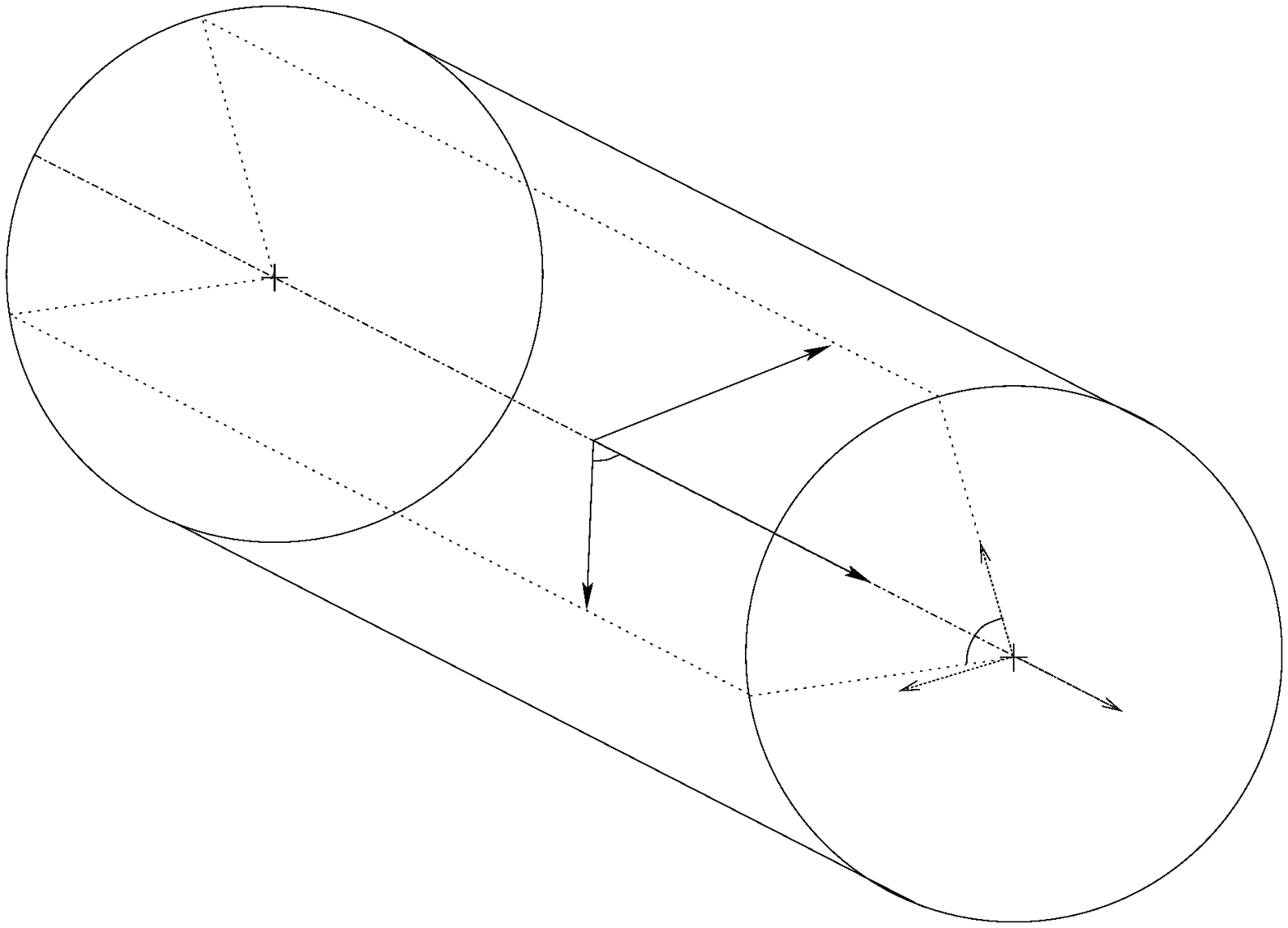}
\put(-73,66){$\tilde{p}_a$}
\put(-118,45){$p_j$}
\put(-86,103){$\tilde{p}_k$}
\put(-111,69){\small{$\theta_j$}}
\put(-58,50){\small{$\phi_j$}}
\put(-43,64){$x$}
\put(-66,33){$y$}
\put(-22,31){$z$}
\end{center}
\caption{Parameterization of the angular variables for the initial-state 
splitting $\tilde{p}_a \to p_a+p_j$. Here $\tilde{p}_k$ is the spectator 
parton selected to define the azimuthal variable $\phi_j$.}
\label{fig:angles_ini}
\end{figure}

In the case of initial-state splitting, the kinematics is simpler and the 
choice of the splitting variables is more straightforward. We consider
\bea
\tilde{p}_a & = & x\,p_a  \\
\tilde{p}_b & = & p_b
\eea
and we take $x$ to be our soft variable, the soft limit being $x\to1$. In the 
frame where the total momentum \textit{before} the splitting, $Q=p_a+p_b$, is 
at rest, the relation between $E_j$ and $x$ is particularly simple:
\beq
E_j = \frac{1}{2}\,\sqrt{\Qsq}\,(1-x) \;.
\eeq
Note that $\Qsq=\tilde{K}^2/x$, where $\tilde{K}=\tilde{p}_a+\tilde{p}_b$. We 
build our reference frame by identifying the $z$-axis with the vector 
$\tilde{p}_a$, and select one spectator parton $\tilde{p}_k$ so that the 
plane $\tilde{p}_a$-$\tilde{p}_k$ defines the azimuthal variable $\phi_j$. 
Finally, the angle between parton $p_j$ and the beam axis defines the 
collinear variable $\theta_j$ as shown in Figure \ref{fig:angles_ini}. 
The integration of the splitting phase space runs in the case of 
initial-state emission over the variables 
\bea
x & \in & [0,1] \;, \\
c \equiv \cos\theta_j & \in & [-1,1] \;, \\
\phi \equiv \phi_j & \in & [0,2\pi] \;.
\eea
The soft and collinear limits correspond to $x\to1$ and $c\to1$ respectively.

\subsection{Semi-numerical approach}
For the calculation of the integrated subtraction terms, we adopt the 
spin-averaged version of the splitting functions described in 
Ref. \cite{Nagy:2008ns}. In $d=4-2\epsilon$ dimensions, one has:
\bea
\overline{W}^{(ii,j)} & = & \,\, F_i \sum_{\bar{s}_i=\pm} 
W^{(ii,j)}_{\tilde{s}_i,\tilde{s_i}} \\
\overline{W}^{(ik,j)} & = & \,\, \frac{1}{2(1-\epsilon)} 
\sum_{\bar{s}_i=\pm} W^{(ik,j)}_{\tilde{s}_i,\tilde{s_i}} \;.
\eea
The average factor $F_i$ takes the value $1/2$ if the splitting parton 
$\tilde{p}_i$ is a quark, and $1/(2(1-\epsilon))$ if it is a gluon.
Including the color operator and exploiting the invariance of the 
matrix element under color rotations,
\beq
\left( \mathbf{T}_i \cdot \mathbf{T}_i \right) = -\sum_{k \ne i}\,\frac{1}{2} 
\left[ \, \left( \mathbf{T}_i \cdot \mathbf{T}_k \right) + 
\left( \mathbf{T}_k \cdot \mathbf{T}_i \right) \, \right] \;,
\eeq 
one can arrange the full integrand for a given mapping 
$\{\tilde{p}\}_{m}^{(ij)}$ in the form
\beq
\frac{1}{2} \left[ \, \left( \mathbf{T}_i \cdot \mathbf{T}_k \right) + 
\left( \mathbf{T}_k \cdot \mathbf{T}_i \right) \, \right]  
\left[ \overline{W}^{(ii,j)} - \overline{W}^{(ik,j)} \right] \;.
\eeq
Following \cite{Nagy:2008ns}, we further split the integrand 
$\left[ \overline{W}^{(ii,j)} - \overline{W}^{(ik,j)} \right]$ into two pieces,
\beq
\label{eq:intdipsplit}
\overline{W}^{(ii,j)} - \overline{W}^{(ik,j)} = 
\underbrace{(\overline{W}^{(ii,j)} - 
\overline{W}^{(ii,j), eik})}_{\mathcal{D}_{ii}} + 
\underbrace{(\overline{W}^{(ii,j), eik} - 
\overline{W}^{(ik,j)})}_{\mathcal{I}_{ik}} \;,
\eeq
where $\overline{W}^{(ii,j), eik}$ is based on the eikonal 
approximation (\ref{eq:eikonal}).
This splitting has the advantage that the first term, $\mathcal{D}_{ii}$, 
can only have a collinear singularity by construction, while the second 
term, $\mathcal{I}_{ik}$, can be both soft- and collinear-divergent. On the 
other hand, the latter can be expressed in the simple universal 
form \cite{Nagy:2008ns}
\beq
\mathcal{I}_{ik} = 4\pi\as \, A_{ik} \frac{-\left( (p_j \cdot p_k)\,p_i 
- (p_i \cdot p_j)\,p_k  \right)^2}{(p_i \cdot p_j)^2 (p_j \cdot p_k)^2} \;.
\eeq

Our goal is to integrate (\ref{eq:intdipsplit}) over the whole phase space 
of the splitting. For clarity, we will present the cases of initial- and 
final-state emission separately. It is convenient to make explicit reference 
to our splitting variables, defined in Section \ref{sec:splitting_variables}, 
and write the integrated subtraction terms as follows: 
\bea
D_{ii} & \equiv & \int d\xi_{fin}^d\,\mathcal{D}_{ii}^d = 
\int de\,dc\,d\phi\,\,J_{fin}^d(e,c,\phi)\,\mathcal{D}_{ii}^d(e,c,\phi) \;, 
\label{eq:intdip1} \\ 
I_{ik} & \equiv & \int d\xi_{fin}^d\,\mathcal{I}_{ik}^d = 
\int de\,dc\,d\phi\,\,J_{fin}^d(e,c,\phi)\,\mathcal{I}_{ik}^d(e,c,\phi) \;, 
\label{eq:intdip2}  \\ 
D_{aa}(\cdot) & \equiv & \int d\xi_{ini}^d\,\mathcal{D}_{aa}^d = 
\int dx\,dc\,d\phi\,\,J_{ini}^d(x,c,\phi)\,\mathcal{D}_{aa}^d(x,c,\phi) \;, 
\label{eq:intdip3}  \\ 
I_{ak}(\cdot) & \equiv & \int d\xi_{ini}^d\,\mathcal{I}_{ak}^d = 
\int dx\,dc\,d\phi\,\,J_{ini}^d(x,c,\phi)\,\mathcal{I}_{ak}^d(x,c,\phi) \;. 
\label{eq:intdip4}
\eea
Here, the superscript $d$ means that the corresponding quantity is evaluated 
in $d$ space-time dimensions, while $J^d$ represents the Jacobian of the 
phase space measure associated with our splitting variables. When the 
splitting involves an initial-state parton, we use the symbol "$(\cdot)$" 
to clarify that the integrated term is a distribution in the $x$ variable, 
thus the corresponding integral is well defined only in a  convolution with 
some test function $f(x)$.

It has been already pointed out that, in consequence of the increased 
complexity of the mapping, a fully analytic evaluation of the 
integrals (\ref{eq:intdip1}), (\ref{eq:intdip2}), (\ref{eq:intdip3}),
 (\ref{eq:intdip4}) is challenging in the Nagy-Soper scheme. A viable 
alternative is to use numerical approaches, such as Gaussian integration 
or Monte Carlo, to integrate over the splitting phase space. Given the 
general complexity of the integrands, we decided to adopt a 
\textit{semi-numerical} approach, namely to consider analytic integration 
when possible, and Monte Carlo integration otherwise.
A crucial observation is that the general dependence of the integrands upon 
the azimuthal variable $\phi$ is simple. In fact, up to 
$\mathcal{O}(\epsilon^0)$, all the azimuthal integrals can be classified 
into three groups:
\bea
& & \int_{0}^{2\pi} d\phi \,\, \frac{1}{x+y \cos\phi}  = 
\frac{2\pi}{\sqrt{x^2-y^2}}\,\mbox{sign}(x) \;, \\
& & \int_{0}^{2\pi} d\phi \,\, \frac{\log(\sin^2\phi)}{x+y \cos\phi}  =  
\frac{4\pi}{\sqrt{x^2-y^2}} \, 
\log\left(\frac{2\sqrt{x^2-y^2}}{2x+\sqrt{x^2-y^2}}\right) \;, \\
& & \int_{0}^{2\pi} d\phi \,\, \log^2(\sin^2\phi)  =  
\frac{2\pi^3}{3} + 2\pi\,\log^2(4) \;.
\eea
Here $x,y$ denote generic real parameters.
The dependence on the soft and collinear variables is not as simple and 
leads to more complicated expressions that deserve a numerical treatment. 
As an example, we will illustrate the case of the integrals 
(\ref{eq:intdip2}) and (\ref{eq:intdip4}) in the case where the emitting 
parton is massless. We highlight these cases because they exhibit some 
complication compared to other integrals, helping us to explain our 
approach in its most general form.

Let us start with the case of final-state interference, Eq.(\ref{eq:intdip2}). 
After the azimuthal variable has been integrated out, one is left with a new 
integrand function $\mathcal{F}_{ik}^d(e,c)$ which is divergent in proximity 
of the soft ($e\to0$) and/or collinear ($c\to1$) limit. Its divergent 
structure is known from QCD and allows us to write,
\beq
\mathcal{F}_{ik}^{d=4-2\epsilon}(e,c) = 
\frac{\mathcal{N}_{ik}^{d=4-2\epsilon}(e,c)}{(1-c)^{1+\epsilon}\,e^{1+2\epsilon}} \;,
\eeq
where the numerator $\mathcal{N}(e,c)^{d=4}_{ik}$ is a regular function in the 
full domain. Using  dimensional regularization, $d=4-2\epsilon$, one converts 
the singular structure that appears implicitly at the integrand level in the 
form of explicit $1/\epsilon$ and $1/\epsilon^2$ poles. To facilitate the pole 
extraction, we add and subtract divergent counterterms to the original 
integral as follows:
\bea
I_{ik} & = & \int de\,dc\,\,\mathcal{F}_{ik}^{d=4-2\epsilon}(e,c) \label 
{eq:subtractionfin} \\
& = & \int de\,dc\, \frac{1}{(1-c)e} \left[ \mathcal{N}_{ik}^{d=4}(e,c) - 
\mathcal{N}_{ik}^{d=4}(0,c) - \mathcal{N}_{ik}^{d=4}(e,1) + 
\mathcal{N}_{ik}^{d=4}(0,1) \right] \label{eq:intdipfin1} \\ 
& + & \int de\,dc\, \frac{1}{(1-c)^{1+\epsilon}\,e^{1+2\epsilon}} 
\left[ \mathcal{N}_{ik}^{d=4-2\epsilon}(0,c) - 
\mathcal{N}_{ik}^{d=4-2\epsilon}(0,1) \right] \label{eq:intdipfin2} \\ 
& + & \int de\,dc\, \frac{1}{(1-c)^{1+\epsilon}\,e^{1+2\epsilon}} 
\left[ \mathcal{N}_{ik}^{d=4-2\epsilon}(e,1) - 
\mathcal{N}_{ik}^{d=4-2\epsilon}(0,1) \right] \label{eq:intdipfin3} \\ 
& + & \int de\,dc\,\, \frac{1}{(1-c)^{1+\epsilon}\,e^{1+2\epsilon}} \, 
\mathcal{N}_{ik}^{d=4-2\epsilon}(0,1) \label{eq:intdipfin4}\,.
\eea
Now the integral (\ref{eq:intdipfin1}) is convergent by construction and 
can be solved numerically in $d=4$ dimensions. The full singular structure 
is entailed in (\ref{eq:intdipfin2}), (\ref{eq:intdipfin3}), 
(\ref{eq:intdipfin4}) and can be directly extracted by expanding the 
integrand in powers of $\varepsilon$. Thus, from (\ref{eq:intdipfin2}) 
and (\ref{eq:intdipfin3}) one gets $1/\epsilon$ poles related to soft and 
collinear singularities, while the soft-collinear $1/\epsilon^2$ pole is a 
genuine contribution of (\ref{eq:intdipfin4}). After expansion, the result 
is cast in the form
\bea
I_{ik} & = & \frac{1}{\epsilon^2}\,G_{ik}^{(2)} + 
\frac{1}{\epsilon}\int de\,dc\,\,\mathcal{G}_{ik}^{(1)}(e,c) + 
\int de\,dc\,\,\mathcal{G}_{ik}^{(0)}(e,c) \\
& = & \frac{1}{\epsilon^2}\,G_{ik}^{(2)} + \frac{1}{\epsilon}\,G_{ik}^{(1)} + 
\int de\,dc\,\,\mathcal{G}_{ik}^{(0)}(e,c) \;.
\eea
Note that this is a two-dimensional application of the well-known relation
\beq
\frac{1}{(1-z)^{1+\epsilon}} = -\frac{1}{\epsilon}\,\delta(1-z) + 
\left( \frac{1}{1-z} \right)_+  + \, \cdots \,\;.
\eeq
The coefficients of the poles are integrals that are simple enough to be 
performed fully analytically, while the finite part is left to Monte Carlo, 
improved with stratified sampling \cite{vanHameren:2007pt} . Both divergent 
and finite pieces are incorporated into the $\mathbf{I}(\varepsilon)$ operator.

Let us now turn to the case of initial-state interference, 
Eq.(\ref{eq:intdip4}). We consider a generic test function $f(x)$ in 
convolution with the integrated subtraction term, and rewrite the latter 
in the standard form
\bea
I_{ak}(f) & = & \int dx\,dc\,\,f(x)\,\mathcal{F}_{ak}^d(x,c) \nonumber \\
& = &  \int dx\,dc\,\left[ f(x)\,\mathcal{F}_{ak}^d(x,c) - 
f(1)\,\mathcal{F}_{ak}^d(x,c) \right] + 
\int dx\,dc\,\,f(1)\,\mathcal{F}_{ak}^d(x,c) \\
& = &  \int dx\,dc\,\, f(x) \left( \mathcal{F}_{ak}^d(x,c) \right)_{+}  + 
f(1)\,\int dx\,dc\,\, \mathcal{F}_{ak}^d(x,c) \\
& \equiv & \int dx\,f(x)\,\left[ \left(U_{ak}^d(x)\right)_{+} + 
\delta(1-x)\,V_{ak}^d \right] \;,
\eea
where $\left(U_{ak}^d(x)\right)_{+}$ and $V_{ak}^d(x)$ are fully 
integrated over the collinear variable $c$:
\bea
\left(U_{ak}^d(x)\right)_{+} & = & \int dc\,\, 
\left( \mathcal{F}_{ak}^d(x,c) \right)_{+} \;, \\
V_{ak} & = & \int dx\,dc\,\,\mathcal{F}_{ak}^d(x,c) \;.
\eea
As in the previous example, one can write
\beq
\mathcal{F}_{ak}^{d=4-2\epsilon}(x,c) = 
\frac{\mathcal{N}_{ak}^{d=4-2\epsilon}(x,c)}{(1-c)^{1+\epsilon}\,
(1-x)^{1+2\epsilon}} \;,
\eeq
where the numerator function is finite in the full domain. The pole structure 
of $\left(U_{ak}^d(x)\right)_{+}$ and $V_{ak}^d(x)$ can now be extracted using 
the outlined subtraction method. As to the first element, only a collinear 
divergence is allowed and the subtraction procedure is simplified:
\bea
\left(U_{ak}^d(x)\right)_{+} & = & \int dc\,\, \left( \frac{1}{(1-c)(1-x)} 
\left[ \mathcal{N}_{ak}^{d=4}(x,c) - \mathcal{N}_{ak}^{d=4}(x,1) \right] 
\right)_{+} \\
& + & \int dc\,\, \left( \frac{1}{(1-c)^{1+\epsilon}\,(1-x)^{1+2\epsilon}} \, 
\mathcal{N}_{ak}^{d=4-2\epsilon}(x,1) \right)_{+}
\eea
In the second case, $V_{ak}$, both collinear and soft singularities are 
allowed and the procedure is the same as the one described 
in Eq.(\ref{eq:subtractionfin}).
Expanding the integrands in powers of $\epsilon$, the results are finally 
cast in the form
\bea
\left(U_{ak}^d(x)\right)_{+} & = & \frac{1}{\epsilon}
 \left( K_{ak}^{(1)}(x) \right)_{+} + \int dc\,\, 
\left( \mathcal{K}_{ak}^{(0)}(x,c) \right)_{+} \;, \\
V_{ak} & = & \frac{1}{\epsilon^2}\,G_{ak}^{(2)} + 
\frac{1}{\epsilon}\,G_{ak}^{(1)} + \int dx \, dc\,\,\mathcal{G}_{ak}^{(0)}(x,c) \;.
\eea
The singular part of $V_{ak}$ is incorporated into the 
$\mathbf{I}(\varepsilon)$ operator. Note that  the single pole of 
$\left(U_{ak}^d(x)\right)_{+}$ is canceled by the corresponding singularity 
arising from  the collinear counterterm $\mathcal{A}^C$, namely it is 
absorbed into a re-definition of the parton distribution functions. The 
finite pieces of both $U_{ak}(x)$ and $V_{ak}$ are all implemented in the 
$\mathbf{KP}$ operator. We stress that splitting an $x$-independent finite 
piece between the $\mathbf{I}$ and the ($x$-dependent) $\mathbf{KP}$ 
operator is a totally arbitrary operation, provided one casts them 
in the form of $\delta(1-x)$ contributions.

%-------------%-------------%-------------%-------------
%

%
%-------------%-------------%-------------%-------------
\section{Implementation in HELAC-DIPOLES}

\textsc{Helac-Dipoles} is a general purpose package  for the evaluation 
of the real emission corrections at next-to-leading order in QCD. Based 
on the framework of \textsc{Helac-Phegas} 
\cite{Kanaki:2000ey,Papadopoulos:2000tt,Kanaki:2000ms,Cafarella:2007pc}, it has 
been originally built using a modified version of the Catani-Seymour dipoles 
which applies to arbitrary helicity eigenstates of the splitting partons 
\cite{Czakon:2009ss}. The package has been  used together with 
\textsc{Helac-1Loop} 
\cite{Ossola:2006us,Mastrolia:2008jb,vanHameren:2009dr,Bevilacqua:2011xh}, 
\textsc{CutTools} \cite{Ossola:2007ax,Ossola:2008xq,Draggiotis:2009yb} 
and \textsc{OneLOop} \cite{vanHameren:2010cp} 
in the  computation of full QCD corrections to several $2\to4$ processes 
at the LHC and the Tevatron \cite{Bevilacqua:2012em,Bevilacqua:2011aa,
Bevilacqua:2010qb,Worek:2011rd,Bevilacqua:2009zn,Bevilacqua:2013taa,
Bevilacqua:2010ve}. 

The structure of the calculation closely follows the general discussion 
of Section \ref{sec:general_framework}. Following 
Eqs.(\ref{eq:subtrreal}-\ref{eq:KPop}), the final output obtainable 
from the code consists of three contributions evaluated 
separately\footnote{We note that evaluating the contributions 
from subtracted real part, $\mathbf{I}$ operator and $\mathbf{KP}$ operator 
in a  single run instead of three separate steps, as proposed 
in \cite{Nagy:1996bz}, may improve the overall efficiency of the 
calculation especially in case of large cancellations between the 
various terms. We leave this optimization for future developments of 
the code.}: the subtracted real part, the $\mathbf{I}$ operator and 
the $\mathbf{KP}$ operator contributions. Although the three terms are 
individually dependent on the subtraction scheme employed, their sum is 
independent and can be used in order to cross-check our implementation 
against Catani-Seymour results.

In the present extension, we have incorporated the new subtraction method 
based on the Nagy-Soper formalism preserving at the same time all the 
optimizations available in the code. For a detailed description of the 
package functionalities, we refer to the existing literature 
\cite{Czakon:2009ss,Bevilacqua:2011xh}. We emphasize that all those 
elements of the calculation that are inherent to subtraction, but not 
dependent on a specific scheme, do not require to be implemented again: 
for example, the Born matrix elements and the color correlators are 
already provided by the framework of \textsc{Helac-Dipoles}. This fact 
dramatically simplifies our implementation. 

The construction of the Nagy-Soper subtraction terms is dictated by the 
form of the splitting functions introduced in Section 
\ref{sec:splittingfunctions}, see for instance 
Eqs.(\ref{eq:splf_qqg})-(\ref{eq:splf_ggg}). An interesting feature 
of these functions is that they contain generic  spinors and polarization 
vectors, which enables them to treat simultaneously fixed helicities as 
well as  random polarization states. For comparison, the Catani-Seymour 
dipoles as to our implementation  \cite{Czakon:2009ss} can only work with 
helicity eigenstates. An extension of the Catani-Seymour formalism for 
randomly polarized partons has been recently presented 
in \cite{Goetz:2012uz} and is characterized by the introduction of 
additional subtraction terms.

The basic idea of the random polarization method is to replace the 
polarization state with a linear combination of helicity eigenstates 
\cite{Draggiotis:1998gr,Draggiotis:2002hm}. 
For example, the polarization vector of a gluon is written as follows
\beq
\varepsilon^\mu(k,\phi) \equiv e^{i\phi}\varepsilon^\mu(k,+) + 
e^{-i\phi}\varepsilon^\mu(k,-) \;,
\eeq
where $\varepsilon^\mu(k,\pm)$ are the helicity eigenstates and 
$\phi \in [0,2\pi]$ is a phase parameter. In this way one can replace 
a discrete sum over the helicities by an integration over the phase 
parameter $\phi$ according to the formula
\beq
\sum_{\lambda} \vert \mathcal{M}_\lambda \vert^2 = \frac{1}{2\pi} 
\int_{0}^{2\pi} d\phi \, \vert \mathcal{M}_\phi \vert^2 \;.
\eeq
In other words, this method extends the sampling phase space with additional
 degrees of freedom represented by the parameters $\phi$'s, one for each 
randomly polarized particle. It should be clear that, in comparison with 
the approach of exact helicity summation, a higher number of points is  
required in order to reach the same integration accuracy. On the other 
hand, the computational complexity of the calculation is decreased in this 
way by a factor $2^{n_2}3^{n_3}$, where $n_2$ and $n_3$ denote the numbers of 
particles with 2 and 3 polarization states. Furthermore, since for every 
value of $\phi$ both helicities contribute, one does not expect large 
differences in the values of $\vert \mathcal{M}_\phi \vert^2$, and the 
resulting quasi-flat distribution in the Monte Carlo sampling should 
lead to an overall satisfactory convergence of the integration.

We provide \textit{random polarization sampling} as a further option 
available for the Nagy-Soper scheme. This is an alternative to the 
existing \textit{random helicity sampling} optimization,  which uses
stratified sampling over the different (incoherent) helicity assignments 
of partons \cite{Czakon:2009ss}. The option for the spin sum treatment 
can be controlled by the user in the  configuration file 
\texttt{dipoles.conf}. A detailed description of this files
 is reported in the Appendix.

Besides random polarization sampling, which is an important speedup in every
calculation, we also provide {\it random sampling over color}, 
or {\it color Monte
  Carlo}, for the subtracted real radiation part. This functionality
provides an important speedup for matrix elements with a large number
of colored external states. We follow the general
ideas from \cite{Papadopoulos:2005ky,Bevilacqua:2009zn}, 
which are also  an essential ingredient of 
\textsc{Helac-1Loop}. Consider a diagram with an arbitrary number
of in- and/or out-going quarks, anti-quarks and gluons. We shall contract
the gluon color index belonging to the adjoint representation with the
generator matrix $T^a_{ij}$. Thus out-going quark and in-going
anti-quark indices transform according to the ${\bf 3}$
representation, out-going anti-quark and in-going quark indices
according to the $\overline{\bf 3}$ representation, whereas gluon
indices according to the ${\bf 3}\otimes \overline{\bf 3}$
representation. Using the well-known identities
\beq
i f^{abc} = \frac{1}{T_F} \mbox{Tr} \left( T^a T^b T^c - T^c T^b T^a
\right) \; ,
\eeq
and
\beq
\label{eq:cvitanovic}
T^a_{ij} T^a_{kl} = T_F \left( \delta_{il}\delta_{kj} - \frac{1}{N_c}
\delta_{ij}\delta_{kl}\right) \; ,
\eeq
one can reduce the contribution of any diagram to the form
\beq
\label{eq:colorflow}
\sum_{\sigma} {\cal A}(\sigma) \delta_{i_1 i_{\sigma(1)}} \dots
\delta_{i_n i_{\sigma(n)}} \; ,
\eeq
where the first index of every delta transforms according to the ${\bf
  3}$ representation and the second according to the $\overline{\bf
  3}$ representation, $\sigma$ is a permutation, and ${\cal
  A}(\sigma)$ the amplitude corresponding to the given
permutation. This representation is called the color flow
representation, because the color ``flows'' from one index to the
other within a link given by a delta function. The square of the
amplitude summed over color is obtained by summing products of deltas
over the indices. Internally, \textsc{Helac} uses the color flow
representation, but because summation over color mixes different
color flows, the color Monte Carlo uses true color configurations. In
fact, a single true color configuration is generated for every phase space
point with a flat distribution. Subsequently, the color flows, which
are compatible with this color configuration are determined. For this,
it is sufficient to check which delta products in
Eq.(\ref{eq:colorflow}) do not vanish for the given assignment of the
color indices. Finally, the amplitudes are evaluated for the
color flows determined in this way. This approach speeds up the
calculation by a large factor dependent on the process. Within a
subtraction scheme, it is important to have subtraction terms, which
match the singularities of the amplitude for a given true color
assignment. A given singular limit involves two partons in the
collinear case, or three partons in the soft case. The colors of the
remaining partons are simply copied from the original
amplitude. During the determination of the color flows for the
(color correlated) amplitude of the subtraction term, one then takes
into account that a parton was obtained by merging two partons. For
this, the color flows have to be generated according
to Eq.~(\ref{eq:cvitanovic}) in order to correspond to a color
correlator ${\bf T}_i \cdot {\bf T}_k$ insertion. Notice that, in case
the splitting pair contains more than one gluon, one has to decide,
which gluon is to be considered soft in the determination of the color
flow (the missing index $j$ from the color correlator). This is done
by taking the gluon with the smaller energy as the soft one. One can
show that such an approach reduces to the usual one (subtraction term
amplitudes summed over color independently of the summation over color
of the original amplitude) upon summation over all colors of the
original amplitude. We have, of course, checked the pointwise
convergence of our implementation for arbitrary configurations.

%
%-------------%-------------%-------------%-------------

\section{Numerical Study}

%-------------%-------------%-------------%-------------
%

%
%-------------%-------------%-------------%-------------

\subsection{Input parameters and phase space cuts}

%-------------%-------------%-------------%-------------

In order to test our implementation and compare it with that of the
Catani-Seymour subtraction scheme, we have to choose some specific
processes. 
Throughout this study we consider proton-proton  collisions at the
LHC with a center-of-mass energy of 8 TeV.  We will concentrate on the
following partonic subprocesses $gg\to t\bar{t}b\bar{b}g$,  $gg\to
t\bar{t}t\bar{t}g$, $gg\to b\bar{b}b\bar{b}g$, $gg\to t\bar{t}ggg$
that give dominant contributions to the subtracted real emissions at
${\cal O}(\alpha_s^5)$  for the corresponding processes  $pp \to
t\bar{t}b\bar{b} +X$, $pp \to t\bar{t}t\bar{t} +X$, $pp \to
b\bar{b}b\bar{b} +X$ and $pp \to t\bar{t}jj +X$. These processes
represent a high level of complexity and test almost all aspects of
the software, as they involve both massive and massless states.
Basic selection cuts
are imposed on jets that are reconstructed via the IR-safe
anti-$k_T$ jet algorithm \cite{Cacciari:2008gp} 
with radius parameter  $R = 1$. They are
reconstructed out of all final-state partons that  are lying in the
pseudo-rapidity range accessible to the LHC detectors, {\it i.e.} within
the range of $|\eta|<5$. All jets are required to carry $p_T(j)> 50$
GeV  and  to be located in the rapidity-range of $|y(j)|<2.5$. They are 
also made to be well separated in the rapidity-azimuthal angle plane with 
 $\Delta R(jj)>1$.  Let us note that $j$
corresponds to a light  jet only, outgoing top and anti-top quarks are
left on-shell, they do not comply to any cut selection.  Let us also add,
that in case of the $gg\to t\bar{t}ggg$ process jets are ordered
according to their $p_T$ and cuts are applied only on the two hardest
jets.  The mass of the top quark is set to $m_t = 173.5$ GeV 
\cite{Beringer:1900zz}, the
bottom quark is considered  to be massless.  Results are presented for
the NLO CT10 parton distribution functions \cite{Lai:2010vv} with 
five active flavors
and the corresponding two-loop  $\alpha_s$.  The renormalization and
factorization scales are set to the scalar sum of the  jet transverse
masses, {\it i.e.}
\begin{equation} 
H_T=\sum m_T(j)\,, 
\end{equation}  
where for the top quark $m_T(t)=\sqrt{m_t^2+p^2_T(t)}$ and for light
jets (also tagged bottom-jets) $m_T(j)=p_T(j)$. A factor of $1/4$ has
been added for all but the $gg\to b\bar{b}b\bar{b}g$  process where
$\mu_R=\mu_F=\mu_0=H_T$ has been chosen instead. If not specified
otherwise full summation over all color configurations is  assumed
together with random helicity sampling.  The phase space integration
is  performed with the help  of the Monte Carlo generator
\textsc{Kaleu} \cite{vanHameren:2010gg} including  a multi-channel
approach \cite{Kleiss:1994qy} for  separate channels that are
associated with the squared real emission matrix element  along with 
each subtraction term \cite{Bevilacqua:2010qb}.   
The integration  over the fractions of the momenta of
the initial partons weighted by the parton distribution functions is
performed with the help of \textsc{Parni} \cite{vanHameren:2007pt}.

%
%-------------%-------------%-------------%-------------

\subsection{Comparisons with the Catani-Seymour scheme}

%-------------%-------------%-------------%-------------
%

We now turn to a comparison of the Nagy-Soper subtraction  scheme
(NS), introduced in the previous sections, with the Catani-Seymour
subtraction scheme (CS), widely used in the calculation of NLO QCD
corrections.  We start our comparison with the total number of
subtraction terms that need to be evaluated in both schemes. They are
presented in  Table \ref{tab:real-emission-dipoles}. The table also 
contains the number of Feynman diagrams corresponding to the subprocesses 
under scrutiny to underline their complexity.  In each case,
five times less terms are needed in the NS subtraction scheme
compared to the CS scheme.  The difference corresponds to  the total
number of possible spectators for a $2\to 5$ process, which are
relevant in the CS case, but not in the NS case.

Real emission cross sections together with their absolute and relative
errors, the latter ones being expressed in $\%$, are given in Table
\ref{tab:real-emission-sigma} and Table \ref{tab:real-emission-error}.
Results are shown for the CS  dipole subtraction, with $(\alpha_{max}
= 0.01)$ and without $(\alpha_{max} = 1)$ restriction  on the phase
space of the subtraction \cite{Frixione:1995ms,Nagy:1998bb,Nagy:2003tz}, 
and the new NS scheme. All results are
obtained  for the same number of  phase space points before cuts. We
have used $64\times 10^{6}$ points for both integrated subtraction terms,
{\bf I}-  and {\bf KP}, while  for the subtracted  real emission part
$50 \times 20\times 10^{6}$  phase-space points have been generated.
For all cross sections the resulting relative errors are well below
$1\%$, the largest one being five per mill, for details see Table
\ref{tab:real-emission-error}.  In addition, the actual maximal difference
between two evaluations of a given cross section is  twice the  sum of
the corresponding
errors. However, on average, calculations coincide within the sum of
two errors.  The absolute errors given in Table
\ref{tab:real-emission-sigma} in the CS case with $\alpha_{max} =
0.01$ are noticeably higher than the corresponding ones for
$\alpha_{max} = 1$, in fact by a factor of $2-4$ depending on the process. 
However, this difference is reduced down to about $1.5$ for the NS
case vs the CS case with $\alpha_{max} = 1$. Thus, we do not observe
dramatical differences between the NS and CS cases
with $\alpha_{max} = 1$. These absolute errors are a consequence of
the size of the errors of the three different contributions to the
cross section. Since we have fixed the number of events for the
single particle phase space integrated contributions ({\bf I} and {\bf
  KP} contributions) they
might have an important effect on the final error. This is indeed the
case for the CS scheme with $\alpha_{max} = 0.01$. As the {\bf I} and {\bf
  KP} contributions are not computationally intensive, a meaningful
comparison of the schemes involves the subtracted real radiation
contribution only. This comparison is performed below.
Let us note, that we do not show any
results for distributions, since such comparisons have already been
performed for some massive and massless cases in \cite{Bevilacqua:2013taa}, 
and we are confident that the study of total cross sections is sufficient
to draw conclusions on agreement and efficiency.

%----------------------------------------------------------
\begin{table}[h!]
\renewcommand{\arraystretch}{1.5}
\begin{center}
  \begin{tabular}{|c|c|c|c|}
\hline
 \textsc{Process}&
\textsc{Nr. of Dipoles}   &
\textsc{Nr. of Subtractions}  & \textsc{Nr. of } \\
&\textsc{Catani-Seymour}& \textsc{Nagy-Soper} 
& \textsc{Feynman Diagrams}\\
\hline
\hline
$gg\to t\bar{t}b\bar{b}g$ & 55 & 11 & 341\\ \hline
$gg\to t\bar{t}t\bar{t}g$ & 30 & 6 &682 \\ \hline 
$gg\to b\bar{b}b\bar{b}g$ & 90 &  18 &  682\\ \hline
$gg\to t\bar{t}ggg$ & 75 &  15 & 1240 \\
\hline
  \end{tabular}
\end{center}
  \caption{\it \label{tab:real-emission-dipoles}  
Number of Catani-Seymour and Nagy-Soper subtraction terms 
for dominant partonic
subprocesses contributing to the subtracted real emission
contributions at ${\cal O}(\alpha_s^5)$ for the $pp \to
t\bar{t}b\bar{b} +X$, $pp \to 
t\bar{t}t\bar{t} +X$, $pp \to  b\bar{b}b\bar{b} +X$ and $pp \to
t\bar{t}jj +X$ processes at the LHC. The number of Feynman 
diagrams corresponding to the subprocesses is given as well.}
 \end{table}
%----------------------------------------------------------
\begin{table}[h!]
\renewcommand{\arraystretch}{1.5}
\begin{center}
  \begin{tabular}{|c|c|c|c|}
\hline
 \textsc{Process}&
$\sigma_{\rm RE}^{\rm CS \, (\alpha_{max}=0.01)}$ [pb] 
& $\sigma_{\rm RE}^{\rm CS \, (\alpha_{max}=1)}$ [pb] &
$\sigma_{\rm RE}^{\rm NS}$ [pb] \\
\hline
\hline
$gg\to t\bar{t}b\bar{b}g$ 
& $(28.43 \pm  0.13)\cdot10^{-3}$ 
& $(28.39 \pm 0.04)\cdot10^{-3}$ 
& $(28.59 \pm 0.06)\cdot10^{-3}$ \\ \hline
$gg\to t\bar{t}t\bar{t}g$ 
& $(17.03 \pm 0.08)\cdot10^{-5}$ 
& $(16.98 \pm 0.02)\cdot10^{-5}$ 
& $(17.01 \pm 0.03)\cdot10^{-5}$\\ \hline 
$gg\to b\bar{b}b\bar{b}g$ 
& $(65.71 \pm 0.30)\cdot10^{-2}$  
& $(66.24 \pm 0.16)\cdot10^{-2}$ 
& $(66.06 \pm 0.22)\cdot10^{-2}$  \\ \hline
$gg\to t\bar{t}ggg$ 
& $(87.91 \pm 0.17)\cdot10^{-1}$
& $(87.96 \pm 0.07)\cdot10^{-1}$
& $(88.16 \pm 0.08)\cdot10^{-1}$\\
\hline
  \end{tabular}
\end{center}
  \caption{\it \label{tab:real-emission-sigma}   Real emission cross
sections for dominant partonic subprocesses contributing to the
subtracted real emissions at ${\cal O}(\alpha_s^5)$ for the $pp \to
t\bar{t}b\bar{b} +X$, $pp \to t\bar{t}t\bar{t} +X$, $pp \to
b\bar{b}b\bar{b} +X$ and $pp \to t\bar{t}jj +X$ processes at the
LHC. Results are shown for two different subtraction schemes, the
Catani-Seymour (CS)  dipole subtraction, with $(\alpha_{max} = 0.01)$
and without $(\alpha_{max} = 1)$ restriction  on the phase space of
the subtraction, and the new Nagy-Soper (NS) scheme, including the
numerical error from the Monte Carlo integration.}
 \end{table}
%----------------------------------------------------------
\begin{table}[h!]
\renewcommand{\arraystretch}{1.5}
\begin{center}
  \begin{tabular}{|c|c|c|c|}
\hline
 \textsc{Process}&
$\varepsilon_{\rm RE}^{\rm CS \, (\alpha_{max}=0.01)}$ [$\%$] 
& $\varepsilon_{\rm RE}^{\rm CS \, (\alpha_{max}=1)}$ [$\%$] &
$\varepsilon_{\rm RE}^{\rm NS}$ [$\%$] \\
\hline\hline
$gg\to t\bar{t}b\bar{b}g$ & 0.47 & 0.16 & 0.22 \\ \hline
$gg\to t\bar{t}t\bar{t}g$ & 0.51 & 0.11 & 0.19 \\ \hline 
$gg\to b\bar{b}b\bar{b}g$ & 0.46 & 0.25 & 0.33 \\ \hline
$gg\to t\bar{t}ggg$       & 0.20 & 0.08 & 0.09      \\
\hline
  \end{tabular}
\end{center}
  \caption{\it \label{tab:real-emission-error}   Relative error in $\%$
on real emission cross sections for dominant partonic subprocesses
contributing to the subtracted real emissions at ${\cal
O}(\alpha_s^5)$ for the $pp \to t\bar{t}b\bar{b} +X$, $pp \to
t\bar{t}t\bar{t} +X$, $pp \to b\bar{b}b\bar{b} +X$ and $pp \to
t\bar{t}jj +X$ processes at the LHC. Results are shown for two
different subtraction schemes, the Catani-Seymour (CS)  dipole
subtraction, with $(\alpha_{max} = 0.01)$ and without $(\alpha_{max} =
1)$ restriction  on the phase space of the subtraction, and the new
Nagy-Soper (NS) scheme.}
 \end{table}
%----------------------------------------------------------
\begin{table}[h!]
\renewcommand{\arraystretch}{1.5}
\begin{center}
  \begin{tabular}{|c|c|c|c|}
\hline
 \textsc{Process}&
$\varepsilon_{\rm SR}^{\rm CS \, (\alpha_{max}=0.01)}$ [pb] 
& $\varepsilon_{\rm SR}^{\rm CS \, (\alpha_{max}=1)}$ [pb] &
$\varepsilon_{\rm SR}^{\rm NS}$ [pb] \\
\hline\hline
$gg\to t\bar{t}b\bar{b}g$ 
& 4.405 $\cdot \,10^{-5}$ 
& 4.108 $\cdot \,10^{-5}$ 
& 5.424 $\cdot \,10^{-5}$\\ \hline
$gg\to t\bar{t}t\bar{t}g$ 
& 1.356 $\cdot \,10^{-7}$  
& 2.298 $\cdot \,10^{-8}$ 
& 2.377 $\cdot \,10^{-8}$ \\ \hline 
$gg\to b\bar{b}b\bar{b}g$ 
& 1.271 $\cdot \,10^{-3}$  
& 1.494 $\cdot \,10^{-3}$ 
& 2.027 $\cdot \,10^{-3}$   \\ \hline
$gg\to t\bar{t}ggg$ 
& 7.560 $\cdot \,10^{-3}$ 
& 2.290 $\cdot \,10^{-3}$ 
& 6.507 $\cdot \,10^{-3}$ 
\\ \hline
  \end{tabular}
\end{center}
  \caption{\it \label{tab:subtracted-error}  Absolute error  for
subtracted real emission cross sections for dominant partonic
subprocesses contributing to the subtracted real emissions at ${\cal
O}(\alpha_s^5)$ for the $pp \to t\bar{t}b\bar{b} +X$, $pp \to
t\bar{t}t\bar{t} +X$, $pp \to b\bar{b}b\bar{b} +X$ and $pp \to
t\bar{t}jj +X$ processes at the LHC. Results are shown for two
different subtraction schemes, the Catani-Seymour (CS)  dipole
subtraction, with $(\alpha_{max} = 0.01)$ and without $(\alpha_{max} =
1)$ restriction  on the phase space of the subtraction, and the new
Nagy-Soper (NS) scheme.}
 \end{table}
%----------------------------------------------------------
\begin{table}[h!]
\renewcommand{\arraystretch}{1.5}
\begin{center}
  \begin{tabular}{|c|c|c|c|}
\hline
 \textsc{Process}&
$t^{\rm CS }$ [msec]
& $t^{\rm NS}$ [msec] &
$t^{\rm RE}$ [msec] \\
\hline\hline
$gg\to t\bar{t}b\bar{b}g$ & $24.8$ & $13.2$ & $6.5$  \\ \hline
$gg\to t\bar{t}t\bar{t}g$ & $35.7$ & $18.5$ & $11.2$  \\ \hline 
$gg\to b\bar{b}b\bar{b}g$ & $26.6$ & $16.2$ & $10.1$  \\ \hline
$gg\to t\bar{t}ggg$       & $214.8$ & $108.2$ & $48.7$  \\
\hline
  \end{tabular}
\end{center}
  \caption{\it \label{tab:subtracted-cpu}  The CPU time needed to
evaluate  the real emission matrix element together with all the
subtraction  terms for one phase space point for two
subtraction schemes, namely  Catani-Seymour, $t^{\rm CS }$ (for
$\alpha_{max}=1$), and Nagy-Soper, $t^{\rm NS}$.  
For comparison, we also give the CPU time
for the pure real emission matrix element calculation, $t^{\rm
RE}$. All numbers have  been obtained on an  Intel 3.40 GHz processor
with the Intel Fortran  compiler  using the option -fast.}
 \end{table}
%----------------------------------------------------------

We now turn our attention to the subtracted real emission
part only. From the computation time  point of view, this contribution is by
far the dominant  piece  of the  complete  NLO calculation and the
only one that requires a computer cluster. In Table
\ref{tab:subtracted-error} we present  absolute errors  of subtracted
real emission cross sections again for both subtraction schemes and
with or without a restriction on the phase space of the subtraction.
Clearly, both
schemes perform similarly and can be employed with  confidence since
final errors are of the same order.  Let us note here however, that
the number of accepted events ({\it i.e.} the efficiency of the phase
space generator) for the three considered cases, CS with $\alpha_{max}=1$,
CS with $\alpha_{max}=0.01$  and NS, was slightly  different. Just to
give a general idea, for the $gg\to t\bar{t}b\bar{b}g$ subprocess
the total number of the phase space points passing the cuts and contributing
to the calculation of the real emission matrix element together with all
the  subtraction terms, was about $12\times 10^6$,  $6\times 10^6$ and
$9\times 10^6$ respectively, without including phase space points  
that have been evaluated in the optimization phase.

Finally, in Table \ref{tab:subtracted-cpu}, the time measured in
milliseconds, needed to evaluate  the real emission matrix element and
the subtraction terms for one phase space point  is given. All times
correspond to $\alpha_{max}=1$. When all subtraction terms are
included, as it is the case for $\alpha_{max}=1$, the CS subtraction
scheme is slower by a factor of about three to four as compared to the
cost of the pure real emission calculation. In case of the NS  scheme,
however, a slowdown of the order of 1.5 to 2 only is observed. We stress
that if a restriction on the  phase space of the subtraction
is applied, corresponding to  $\alpha_{max}\ll 1$, the cost of the subtracted
real emission part is of the order of the cost of the real emission
itself, {\it i.e} higher only by about $10\%-15\%$.  We have checked
this for the CS case, where this restriction is implemented, assuming
$\alpha_{max}=0.01$. In that case, the total time for one phase
space point has been estimated as an average from about 200 different
phase space points, because for  $\alpha_{max}\ne 1$ a different
number of subtraction terms is evaluated for each phase space point. 

Overall, both schemes, with their different momentum mappings and
subtraction terms, have similar performance and give the same  results
for total real emission  cross sections.

%-------------%-------------%-------------%------------- 
%
\subsection{Random color and polarization sampling }
%
%-------------%-------------%-------------%-------------

In the last part on numerical results, we study the overall
performance of Monte Carlo sampling over color and polarization.  At the very
beginning, real emission cross sections with absolute errors using
random color sampling for 
two different subtraction schemes, the Catani-Seymour (CS)  dipole
subtraction, with $(\alpha_{max} = 0.01)$ and without $(\alpha_{max} =
1)$ restriction  on the phase space of the subtraction, and the new
Nagy-Soper (NS) scheme are given in  Table
\ref{tab:real-emission-sigma-mc}.  Relative errors expressed in $\%$  are
shown in Table \ref{tab:real-emission-error-mc}. We observe
agreement with results presented in Table
\ref{tab:real-emission-sigma} where summation over all color flows
has been performed. However, in the present case the absolute errors are
$3-4$ times higher, which is typical when Monte Carlo sampling is
employed instead of full summation.  Generally speaking, in order to
obtain the same absolute errors $9-16$ times more events would be
required. In the case of  $gg\to t\bar{t}b\bar{b}g$,  $gg\to
t\bar{t}t\bar{t}g$ and  $gg\to b\bar{b}b\bar{b}g$ processes, which are
the same from the color point of view,  the total number of color
flows is 120. Out of those only 98 give non-zero contributions and
are evaluated for each phase space point for full summation
over color. When Monte Carlo sampling is employed instead, the
average number of color flows corresponding to a random color
configuration is only 5. This gives a
speed up  of the order of 20 per phase space point, which is almost
fully absorbed by  the higher statistics that is required to obtain the
same absolute error.  Therefore, we conclude that for processes
dominated by quarks random color sampling performs similarly to full color
summation.  
%
%----------------------------------------------------------
\begin{table}[t!]
\renewcommand{\arraystretch}{1.5}
\begin{center}
  \begin{tabular}{|c|c|c|c|}
\hline
 \textsc{Process}&
$\sigma_{\rm RE, \,COL}^{\rm CS \, (\alpha_{max}=0.01)}$ [pb] 
& $\sigma_{\rm RE, \,COL}^{\rm CS \, (\alpha_{max}=1)}$ [pb] &
$\sigma_{\rm RE, \,COL}^{\rm NS}$ [pb] \\
\hline
\hline
$gg\to t\bar{t}b\bar{b}g$ 
& $(28.91\pm 0.32)\cdot 10^{-3}$ 
& $(28.35\pm 0.14)\cdot 10^{-3}$ 
& $(28.77\pm 0.14)\cdot 10^{-3}$  \\ \hline
$gg\to t\bar{t}t\bar{t}g$ 
& $(16.99\pm 0.10)\cdot 10^{-5}$ 
& $(17.00\pm 0.03)\cdot 10^{-5}$ 
& $(17.01\pm 0.04)\cdot 10^{-5}$ 
\\ \hline 
$gg\to b\bar{b}b\bar{b}g$ 
& $(67.01\pm 0.64)\cdot 10^{-2}$
& $(65.71\pm 0.50)\cdot 10^{-2}$
& $(67.00\pm 0.66)\cdot 10^{-2}$
\\ \hline
$gg\to t\bar{t}ggg$ 
&  $(88.05\pm 0.45)\cdot 10^{-1}$
&  $(88.04\pm 0.37)\cdot 10^{-1}$
& $(87.76\pm 0.31)\cdot 10^{-1}$
\\ \hline
  \end{tabular}
\end{center}
  \caption{\it \label{tab:real-emission-sigma-mc}   Real emission cross
sections for dominant partonic subprocesses contributing to the
subtracted real emissions at ${\cal O}(\alpha_s^5)$ for the $pp \to
t\bar{t}b\bar{b} +X$, $pp \to t\bar{t}t\bar{t} +X$, $pp \to
b\bar{b}b\bar{b} +X$ and $pp \to t\bar{t}jj +X$ processes at the
LHC. Results are shown for random color sampling for 
two different subtraction schemes, the
Catani-Seymour (CS)  dipole subtraction, with $(\alpha_{max} = 0.01)$
and without $(\alpha_{max} = 1)$ restriction  on the phase space of
the subtraction, and the new Nagy-Soper (NS) scheme, including the
numerical error from the Monte Carlo integration.}
 \end{table}
%----------------------------------------------------------
\begin{table}[t!]
\renewcommand{\arraystretch}{1.5}
\begin{center}
  \begin{tabular}{|c|c|c|c|}
\hline
 \textsc{Process}&
$\varepsilon_{\rm RE, \,COL}^{\rm CS \, (\alpha_{max}=0.01)}$ [$\%$] 
& $\varepsilon_{\rm RE, \, COL}^{\rm CS \, (\alpha_{max}=1)}$ [$\%$] &
$\varepsilon_{\rm RE, \, COL}^{\rm NS}$ [$\%$] \\
\hline\hline
$gg\to t\bar{t}b\bar{b}g$ & 1.10 & 0.50 &  0.47 \\ \hline
$gg\to t\bar{t}t\bar{t}g$ & 0.60  & 0.18  &  0.23\\ \hline 
$gg\to b\bar{b}b\bar{b}g$ & 0.96 & 0.76 &  0.98 \\ \hline
$gg\to t\bar{t}ggg$       & 0.52 & 0.42  &  0.35     \\
\hline
  \end{tabular}
\end{center}
  \caption{\it \label{tab:real-emission-error-mc}   Relative error in $\%$
on real emission cross sections for dominant partonic subprocesses
contributing to the subtracted real emissions at ${\cal
O}(\alpha_s^5)$ for the $pp \to t\bar{t}b\bar{b} +X$, $pp \to
t\bar{t}t\bar{t} +X$, $pp \to b\bar{b}b\bar{b} +X$ and $pp \to
t\bar{t}jj +X$ processes at the LHC. Results are shown for 
random color sampling for two
different subtraction schemes, the Catani-Seymour (CS)  dipole
subtraction, with $(\alpha_{max} = 0.01)$ and without $(\alpha_{max} =
1)$ restriction  on the phase space of the subtraction, and the new
Nagy-Soper (NS) scheme.}
 \end{table}
%----------------------------------------------------------
\begin{table}[h!]
\renewcommand{\arraystretch}{1.5}
\begin{center}
  \begin{tabular}{|c|c|c|}
\hline
 \textsc{Process}&
$\sigma_{\rm RE, \,POL}^{\rm NS}$ [pb]  
& $\varepsilon_{\rm RE, \,POL}^{\rm NS}$ [\%]\\
\hline
\hline
$gg\to t\bar{t}b\bar{b}g$ 
& $(28.50\pm 0.06)\cdot 10^{-3}$ &  0.21 \\ \hline
$gg\to t\bar{t}t\bar{t}g$ 
& $(17.01 \pm 0.03)\cdot10^{-5}$ &  0.19 \\ \hline 
$gg\to b\bar{b}b\bar{b}g$ 
& $(66.23 \pm 0.20)\cdot10^{-2}$  & 0.30\\ \hline
$gg\to t\bar{t}ggg$ 
& $(88.16 \pm 0.07)\cdot10^{-1}$ &  0.08\\ \hline
  \end{tabular}
\end{center}
  \caption{\it \label{tab:real-emission-sigma-pol}   Real emission cross
sections for dominant partonic subprocesses contributing to the
subtracted real emissions at ${\cal O}(\alpha_s^5)$ for the $pp \to
t\bar{t}b\bar{b} +X$, $pp \to t\bar{t}t\bar{t} +X$, $pp \to
b\bar{b}b\bar{b} +X$ and $pp \to t\bar{t}jj +X$ processes at the
LHC. Results are shown for random polarization sampling  for 
the new Nagy-Soper (NS) subtraction scheme, including the
numerical error from the Monte Carlo integration. Also given are 
relative errors in $\%$.}
 \end{table}
%----------------------------------------------------------
%
On the other hand, in case of the  $gg\to t\bar{t}ggg$ process, where
the number of  gluons is much higher, the number of all, non-zero and
average color flows per phase space point  is   720, 326 and 6
respectively. Here the  speed up per phase space point is of the order
of 50 and therefore still clearly visible even if  a higher number of
events is generated.  Therefore, we draw the conclusion that random color
sampling is a powerful approach mostly for processes where the number of
gluons is higher and   exceeds the number of quarks.

Finally, in Table \ref{tab:real-emission-sigma-pol} real emission
cross sections for random polarization sampling  for  the new
Nagy-Soper (NS) subtraction scheme, including numerical errors from
the Monte Carlo integration are shown. Also given there are their
relative errors in $\%$. When comparing the numbers presented with
results from Table \ref{tab:real-emission-sigma} and Table
\ref{tab:real-emission-error}  (last column in both cases) where we
make use of random helicity sampling,  not only a perfect agreement
can be noticed but also the same absolute  and relative errors can be
found. In addition, the CPU time that is needed for the evaluation of
one phase space point  is similar in both cases, which clearly
tells us that for the processes under consideration both methods
behave similarly and can be used interchangeably. Moreover, these
two different approaches  can be  utilized to test and compare results. 
%
%-------------%-------------%-------------%-------------

\section{Conclusions and Outlook}

In this paper, we have presented the details of a complete
implementation of the Nagy-Soper subtraction scheme. The most
important aspect was the integration of the subtraction terms over the
one-particle unresolved phase space. We have achieved this by using a
semi-numerical approach, where the azimuthal angle is integrated
analytically, while the remaining angle and energy are integrated
numerically after suitable subtraction. We have implemented our
approach within the \textsc{Helac-Dipoles} framework making use of the
existing structure designed for the Catani-Seymour dipole subtraction
scheme. Thanks to the structure of the splitting functions in the
Nagy-Soper scheme, which are given in terms of actual spinors and
polarization vectors, we were able to provide random polarization
Monte Carlo summation over spins instead of full summation or random helicity
sampling, which is used in the Catani-Seymour case. Additionally, we
have implemented Monte Carlo summation over color by generating random
true color assignments and translating them into the color flow
language. This last optimization will be most useful in large
calculations for many parton final states.

Having two different subtraction schemes available within the same
software allowed us to make efficiency comparisons. There are two
aspects, when comparing Catani-Seymour and Nagy-Soper subtraction. The
first aspect is the evaluation time needed to obtain the contribution
of the real radiation and the subtraction terms for a given phase
space point. By design, the Nagy-Soper scheme has less kinematical
mappings and should therefore be faster. This is indeed what we could
observe. In our applications with a moderate number of final state
partons (five), the differences are less than a factor of two in favor
of Nagy-Soper subtraction as long as there is no restriction on the
dipole phase space in the Catani-Seymour scheme. With a low cutoff in
the latter scheme, both approaches are very similar as far as time is
concerned. The second aspect is the convergence rate of the
integration. We used the same phase space generator, \textsc{Kaleu} in
our comparisons and assumed the same number of generated points. We
observed that the absolute error of the most costly (in terms of
computational time) contribution was slightly worse for Nagy-Soper
than for Catani-Seymour without phase space restriction. In the end,
we are forced to conclude that both schemes are similar in
efficiency. We do not consider differences below a factor
of two in error or time (which are moreover process dependent) a
reason to prefer either scheme.

There are two advantages of our implementation. The first is that we
can now perform better tests in applications by computing real
radiation in two different schemes. The second is that the integrated
subtraction terms  can be used to match the fixed order calculation
onto the Nagy-Soper parton shower. This will be the subject of our
future work. Another direction of research is the implementation of
the phase space cutoffs on the subtraction terms in the Nagy-Soper
scheme. We would also like to investigate, whether having such a
restriction and combining the evaluation of subtracted real radiation
and integrated subtraction terms on a point-by-point basis could
improve the convergence of the calculation.

Finally, let us stress that the software developed in the course of
this work is publicly available\footnote{\tt 
http://helac-phegas.web.cern.ch/helac-phegas/helac-dipoles.html.}.

%-------------%-------------%-------------%-------------
%

\section*{Acknowledgments}

%
%-------------%-------------%-------------%-------------

We acknowledge useful discussions with M.Kr\"amer and Z.Nagy. 

This research was supported by the German Research Foundation (DFG)
via the Sonderforschungsbereich/Transregio SFB/TR-9 {\it
  ``Computational Particle Physics''}. M.C. was also supported by the
Heisenberg programme of the DFG. M.W. acknowledges support by the DFG
under Grant No. WO 1900/1-1 ({\it "Signals and Backgrounds Beyond
  Leading Order. Phenomenological studies for the LHC"}).

%-------------%-------------%-------------%-------------
%

%
%-------------%-------------%-------------%-------------

\appendix
\section{Configuration of \textsc{HELAC-DIPOLES}}

In this Appendix we provide an updated description of the 
\texttt{dipoles.conf} file, which is the configuration file where the user 
can set specific parameters for the calculation of the subtracted real part.
 For a more complete description of the program setup  we refer 
to \cite{Bevilacqua:2011xh}. 
\begin{itemize}
\item \texttt{dipoletype}:  type of the subtraction scheme: 
0 - Catani-Seymour, 1 - Nagy-Soper.
\item \texttt{onlyreal}: if set to true, only real emission, without 
subtraction terms, is calculated. The cuts are specified in {\tt cuts.h}. The 
result must coincide with the one of the original \textsc{Helac-Phegas}
with the same input parameters. This option is included for testing purposes.
\item \texttt{onlylast}: if set to true, only those dipoles will be included, 
which contain the last particle (for correctness it must be a parton). This 
is useful for some processes, where it is clear
that only the last particle can be soft/collinear, and the bookkeeping 
remains simple to obtain the full result at NLO.
\item \texttt{onlydiv}: if set to true, only divergent dipoles will be 
included. Non-divergent dipoles correspond to a pair of massive quarks 
in the final state. They are only useful to get rid of large Sudakov 
logarithms, but are not essential for the finiteness of the real radiation 
contribution. For the Nagy-Soper scheme, non-divergent subtractions are 
not available in the present version.
\item \texttt{hybrid}: if set to true, non-parton polarizations will be 
summed over by a continuous Monte Carlo integration over a phase parameter.
\item \texttt{signmode}: defines how positive and negative contributions 
are to be treated: 0 - the result is left unchanged, both positive and 
negative results are included, 1 - only positive numbers, a negative 
result is set to zero, 2 - only negative numbers, but with the changed sign,  
positive results are set to zero.
\item \texttt{sumtype}:  in the first phase (preferably during the phase-space 
optimization), the summation over helicities of the partons can be performed 
in three different ways: 0 - exact fast summation
(independently for real radiation and dipoles), 1 - exact slow 
summation (for a given helicity configuration both real radiation and 
the dipole sum will be calculated), 2 - Monte Carlo summation over all 
non-vanishing helicity configurations with multichannel optimization. In 
practice, option 2 is recommended. For the Nagy-Soper scheme  
an additional option is available: 3 - random polarization sampling. In
this case, sampling is used throughout the calculation, {\it i.e.} also
during the phase space optimization.
\item \texttt{nsumpol}: number of accepted points to be summed over helicity 
with the method specified by sumtype. The counting starts after phase 
space optimization is finished.
\item \texttt{noptpol}: number of accepted points to be used for helicity 
sampling optimization. During helicity sampling optimization, slow summation 
over helicity configurations (in the sense
defined in the description of sumtype) is performed. It is therefore 
recommended to keep this number relatively small (of the order of a few 
hundred to a thousand).
\item \texttt{nuptpol}: number of accepted points after which an update of 
the helicity sampling weights is performed. This number should be rather 
large for best results (at least an order of
magnitude larger than noptpol).
\item \texttt{alphaMinCut}: lowest value of $\alpha_{min}$, below which a point 
will be rejected altogether, because of the risk of numerical instabilities. 
For the exact definition of $\alpha_{min}$, see {\it e.g.} \cite{Czakon:2009ss}.
\item \texttt{alphaMaxII, alphaMaxIF, alphaMaxFI, alphaMaxFF, kappa}: 
parameters for a restriction on the phase space of the subtraction 
in the Catani-Seymour scheme (see \cite{Czakon:2009ss}). 
For the Nagy-Soper scheme, no 
phase-space restriction is available in the present version.
\item \texttt{colorsampling}: if set to true, Monte Carlo color 
sampling is performed.
\item \texttt{jet algorithm}: type of jet algorithm: 
1 - $k_T$, -1 - anti-$k_T$, 0 - Cambridge/Aachen, see {\it e.g.} 
\cite{Salam:2009jx}.
\item \texttt{number of $b$-jets}: number of tagged bottom-jets in the 
final state.
\item \texttt{max. pseudorapidity of clustered partons, jet resolution 
parameter}: $\eta_{max}$ and $R$, parameters that shape the jet algorithm.
\item \texttt{jetveto}: if set to true, additional jet radiation is vetoed.
\item \texttt{ptveto}: maximum allowed $p_T$ for the extra jet, in case a
jet veto is active.
\end{itemize}

%-------------%-------------%-------------%-------------
%

%%%%%%%%%%%%%%%%%%%%%%%%%%%%%%%%%%%%%%%%%%%%%%%%

%\bibliographystyle{JHEP-2}
%\bibliography{DIPOLES}

\providecommand{\href}[2]{#2}\begingroup\raggedright\endgroup

\end{document}